# Chiral transport of pseudo-spinors induced by synthetic gravitational field in photonic Weyl metamaterials

Hongwei Jia,[1,2] Ruo-Yang Zhang,[1] Wenlong Gao,[3] Shuang Zhang,[4,]* C. T. Chan[1,2,]†

[1]*Department of Physics, the Hong Kong University of Science and Technology, Clear Water Bay, Kowloon, Hong Kong, China.*
[2]*Institute for Advanced Study, the Hong Kong University of Science and Technology, Clear Water Bay, Kowloon, Hong Kong, China.*
[3]*Department of Physics, Paderborn University, Warburger Straße 100, 33098 Paderborn, Germany.*
[4]*School of Physics and Astronomy, University of Birmingham, Birmingham B15 2TT, UK.*
Corresponding authors: *s.zhang@bham.ac.uk, †phchan@ust.hk

Abstract: Weyl particles exhibit chiral transport property under external curved space-time geometry. This effect is called chiral gravitational effect, which plays an important role in quantum field theory. However, the absence of real Weyl particles in nature hinders the observation of such interesting phenomena. In this paper, we show that chiral gravitational effect can be manifested in Weyl metamaterials with spatially controlled nonlocality. This inhomogeneous modulation results in a spatially dependent group velocity in the Weyl cone dispersion, which is equivalent to introducing a curved background space-time (or gravitational field) for Weyl pseudo-spinors. The synthetic gravitational field leads to the quantization of energy levels, including chiral zeroth order energy modes (or simply chiral zero modes) that determine the chiral transport property of pseudo-spinors. The inhomogeneous Weyl metamaterial provides an experimentally realizable platform for investigating the interaction between Weyl particles and gravitational field, allowing for observation of chiral gravitational effect in table-top experiments.

Introduction: The Weyl particles, also called Weyl spinors, are spin 1/2 massless relativistic particles [1] and can be described with Weyl equations. Owing to the chirality of Weyl particles, chiral channels can emerge under an external gauge field, resulting in the chiral transport of the Weyl particles, which is called the chiral gauge effect in particle physics [1-4]. However, the absence of Weyl particles as elementary particles in nature hinders the observation of these interesting chiral gauge effects. The recent discoveries of topological phases in condensed matter and photonics facilitate the observations of many nontrivial topological properties [5-18]. Among various topological materials, systems with Weyl degeneracies between two energy bands have been extensively studied [6-8,11,13-15]. Besides the broadly investigated Fermi-arc surface states protected by non-zero Chern numbers as aconsequence of the bulk-surface correspondence, the transport properties of quasi-particles near the Weyl nodes inside the bulk are also attracting growing attention. Very similar to the real Weyl particles, these properties can also be described by the Weyl equations [11,15,17]. Hence, these quasi-particles are called pseudo-spinors, and such systems provide ideal platforms for the investigation of the intriguing transport properties of Weyl particles, such as the chiral gauge effect.

Weyl degeneracy is highly robust in three-dimensional momentum space because of the codimension 3=3-0 [8,15,19]. It cannot be lifted unless two Weyl nodes with opposite chiralities annihilate each other. Perturbations induced by geometric deformations in crystals or metamaterials can only result in either a position shift of the degeneracy point in momentum space or a change of group velocities near the Weyl cone [19-25]. The position shift corresponds to an artificial U(1) gauge potential or equivalently a synthetic magnetic field that can be experienced by pseudo-spinors [19-26]. Chiral zeroth Landau levels that are propagative in the chiral direction (determined by the chirality of Weyl point and the direction of magnetic field) can be induced by the artificial magnetic field [19-25,27,28], which is called the chiral magnetic effect and have been observed experimentally [22,23]. However, the effects arising from the modulation of group velocities near the Weyl cone have received less attention. Since the group velocities can be interpreted as the space-time metric for the pseudo-spinor [2,19,24,29-36], a curved space-time interpretation (or gravitational field) of the emergent physical effects is highly desirable, and will require comprehensive explorations in theory and experiment.

Chiral gravitational effect is a very important phenomenon in quantum field theory and supergravity theory, which describes the chiral transport properties of Weyl particles when coupled with an external curved space-time geometry background [2,37-39]. It would be extremely difficult to observe chiral gravitational effect in nature due to the absence of Weyl particles as fundamental particles and absence of strongly curved space-time geometries in experimentally accessible conditions. However, it is feasible to realize these effects experimentally in inhomogeneous Weyl materials with spatially controlled group velocities of the Weyl cones (where the group velocity matrix determines the metric tensor of an effective curved space-time [19,24,29-36,40] and hence is akin to artificial gravitational field). Recently, the non-zero chiral anomaly coefficient, which is a signature of gravitational anomaly, was observed experimentally in thermally gradient Weyl semimetals [41] in a flat space-time geometry. However, the formation of chiral channels in inhomogeneous Weyl systems with artificial gravitational field remains unexplored. Thus the elucidation of the physical origin of chiral channels, which are key signatures of the chiral gravitational effect [38,39], will provide guidance for experimental observation.

Here, we theoretically investigate the transport behaviors of pseudo-spinors in curved space-time geometry background via inhomogeneous photonic Weyl metamaterials [7,22]. By spatially tuning the effective constitutive parameters of the metamaterial, the *Weyl cone* dispersion can be tuned locally, resulting in a synthetic frame (and coframe) field that is spatially varying (see FIG. 1(a)-(b), two curved space-time generated by different *Weyl cone* variations in space), which can be interpreted as a gravitational field according to the general relativity theory [1,2]. The Weyl equation describing the Weyl cone is rewritten in the covariant form as required by the parallel transport of pseudo-spinor, with which a Riemannian curvature can be derived from the Yang-Mills gauge theory, serving as the physical origin of the chiral channels. The chiral channels arise from the zeroth order energy modes of the quantized energy levels of pseudo-spinors in the synthetic gravitational background, which can be solved from the covariant form of Weyl equation. The zeroth order modes govern the chiral transport properties of the particles near the band crossings (as displayed in FIG. 1(a)-(b), different curved space-time geometries lead to counter-propagating chiral modes with the presence of space-time curvature). Our work serves as the first realistic proposal for observing the chiral gravitational effect in photonic Weyl systems. It will contribute to the verification of the theories of the interaction between Weyl particles and gravitational fields.

## I. Type I to II transition of Weyl cone due to nonlocality.

We start from the effective media description of a bianisotropic media as a consequence of the homogenization of a Weyl metamaterial, and the constitutive parameters have the following form,

$$\boldsymbol{\varepsilon} = \begin{bmatrix} 1+\dfrac{f_1}{\omega_0^2-\omega^2} & 0 & 0 \\ 0 & 1+\dfrac{f_1}{\omega_0^2-\omega^2} & 0 \\ 0 & 0 & 1 \end{bmatrix}, \boldsymbol{\mu} = \begin{bmatrix} 1+\dfrac{f_2\omega^2}{\omega_0^2-\omega^2} & 0 & 0 \\ 0 & 1+\dfrac{f_2\omega^2}{\omega_0^2-\omega^2} & 0 \\ 0 & 0 & 1 \end{bmatrix}, \boldsymbol{\gamma} = \begin{bmatrix} -\dfrac{\sqrt{f_1 f_2}\,\omega}{\omega_0^2-\omega^2} & 0 & 0 \\ 0 & \dfrac{\sqrt{f_1 f_2}\,\omega}{\omega_0^2-\omega^2} & 0 \\ 0 & 0 & 0 \end{bmatrix}, \quad (1)$$

where $f_1$ and $f_2$ are two real and positive constants (so that the dispersion is consistent with causality), and $\omega_0$ denotes the effective plasma frequency, which can be spatially engineered to be dependent on both position and momentum through structure parameters. The chirality tensor $\boldsymbol{\gamma}$ breaks the parity inversion symmetry, while the system is time reversal invariant. The electric displacement **D** and magnetic inductance **B** can be related to the electromagnetic field (**E** and **H**) via constitutive relations,

$$\begin{aligned} \mathbf{D} &= \varepsilon_0 \mathbf{E} + \mathbf{P}, \quad \mathbf{B} = \mu_0 \mathbf{H} + \mathbf{M} \\ \mathbf{D} &= \varepsilon_0 \varepsilon \mathbf{E} + i\gamma \mathbf{H}/c, \quad \mathbf{B} = \mu_0 \mu \mathbf{H} - i\gamma \mathbf{E}/c \end{aligned} \quad (2)$$

where **P** and **M** are polarization and magnetization fields, respectively. The source-free Maxwell equations ($\nabla \times \mathbf{E} = i\omega \mathbf{B}$, $\nabla \times \mathbf{H} = -i\omega \mathbf{D}$) and the constitutive relations can be transformed into an eigen value problem [42]

$$H\tilde{\psi} = \omega \tilde{\psi}, \quad \text{with } H = N^{-1/2} M N^{-1/2} \text{ and } \tilde{\psi} = N^{1/2} \psi. \quad (3)$$

where the matrix forms $M$, $N$ and $\psi$ are given by

$$M = \begin{bmatrix} 0 & K & 0 & iT_1 \\ -K & 0 & 0 & 0 \\ 0 & 0 & 0 & iT_2 \\ -iT_1^T & 0 & -iT_2 & 0 \end{bmatrix}, \quad N = \begin{bmatrix} I_3 & 0 & 0 & 0 \\ 0 & I_3 & 0 & T_1' \\ 0 & 0 & I_2 \omega_0^2 / f_1 & 0 \\ 0 & T_1'^T & 0 & f_2 I_2 \end{bmatrix}, \quad \psi = \begin{bmatrix} \sqrt{\varepsilon_0} \mathbf{E} \\ \sqrt{\mu_0} \mathbf{H} \\ \mathbf{P}/\sqrt{\varepsilon_0} \\ \mathbf{M}/\sqrt{\mu_0} \end{bmatrix}, \quad (4)$$

The blocks in the matrices ($K$, $T_1$, $T_2$ and $T'_1$) are specified in Eq. (A5), and $I$ is an identity matrix with the subscript 2 or 3 being the dimension. The derivation details of the Hamiltonian from Maxwell equations and constitutive parameters are shown in Appendix A. It is found that $M$ is Hermitian, and $N$ is a positive definite (because $f_1$ and $f_2$ are positive) symmetric matrix, and thus the Hamiltonian $H$ is Hermitian. By solving the plane wave solutions of the Hamiltonian, the dispersion relation can be obtained, as plotted in FIG. 2(a). For the metamaterials described by Eq. (1), there are four Weyl points ($Q1$-$Q4$) located on $k_x$ and $k_y$ axes at angular frequency 11.27rad/ns. Each Weyl point arises from the band crossing between a transversal mode and a longitudinal mode. In the absence of nonlocality, the longitudinal mode is dispersionless, resulting in Weyl degeneracies right at the transition between Type-I and Type-II (FIG. 2c). Non-locality (i.e. $\omega_0$ depends on **k**) can be tailored to tune the dispersion of longitudinal mode to have a positive or negative slope (see FIG. 2b, d) [43], leading to Type-I or Type-II Weyl cones. The relation $\varepsilon_{xx} = \varepsilon_{yy}$, $\mu_{xx} = \mu_{yy}$ and $\gamma_{xy} = \gamma_{yx} = 0$ implies the presence of two mirror symmetry planes in the diagonal directions with respect to *x* and *y* axis. To generate curved space-time geometry, one can spatially vary the group velocities of the Weyl cone. This can be achieved by elaborately designing the **k** dependence of $\omega_0$ at different locations, thus realizing the fine control of the adiabatic variation of the dispersion relation for the longitudinal mode everywhere in the system.

## II. Curved space-time interpretation.

To investigate the transport properties of quasi-particles near the Weyl cone in the inhomogeneous system, we first derive the Weyl equation describing the two bands near the band crossing for a homogeneous system. Taking one of the Weyl points *Q1* with a positive chirality as an example, based on the *k.p* perturbation theory [14,15,17] (see details in Appendix A), the Weyl equation can be derived as,

$$(\bar{\sigma}^0 \partial_t - w\bar{\sigma}^0 \partial_x + v_x \bar{\sigma}^1 \partial_x + v_y \bar{\sigma}^2 \partial_y + v_z \bar{\sigma}^3 \partial_z)\varphi = 0, \tag{5}$$

where $v_x$, $v_y$ and $v_z$ are group velocities in different directions, and $w$ is the velocity that tilts the Weyl cone along $k_x$ direction. $\bar{\sigma}^a$ (*a*=0,1,2,3) in the Weyl equation are defined as $\bar{\sigma}^0 = -\sigma^0$, $\bar{\sigma}^a = \sigma^a$ (*a*=1,2,3), where $\sigma^a$ (*a*=1,2,3) denote the Pauli matrices and $\sigma^0$ is the 2 × 2 identity matrix. Each Weyl point (*Q1-Q4*) has a partner with an opposite chirality (*Q1'-Q4'*) (FIG. 6), which are located in the negative frequency domain (with opposite **k**). Here we focus on *Q1'*, which is the partner of *Q1* and can be described as

$$(\sigma^0 \partial_t - w\sigma^0 \partial_x + v_x \sigma^1 \partial_x + v_y \sigma^2 \partial_y + v_z \sigma^3 \partial_z)\chi = 0. \tag{6}$$

The eigen states $\varphi$ and $\chi$ of Eqs. 5 and 6 are both two-component chiral pseudo-spinors. They are associated with the eigenstates of the full Hamiltonian with a matrix $C$ having elements $C_{i,j} = \langle \tilde{\psi}_j^W | \tilde{\psi}_i \rangle$, and the components of $\varphi$ and $\chi$ can be obtained by picking $C_{i,j}$ that charicaterizes the energy bands forming degeneracies ($\tilde{\psi}_j^W$ and $\tilde{\psi}_i$ are eigenstates at and near the degeneracy point respectively, see details Appendix A). Now we are ready to construct a Dirac equation with Eqs. (5-6),

$$(\gamma^0 \partial_t - w\gamma^0 \partial_x + v_x \gamma^1 \partial_x + v_y \gamma^2 \partial_y + v_z \gamma^3 \partial_z)\phi = 0, \tag{7}$$

where $\phi$ is the four-component Dirac pseudo-spinor $\phi = [\varphi, \chi]^T$, and $\gamma^a$ ($a$=0,1,2,3) are the Dirac matrices in the Weyl representation [1]

$$\gamma^a = \begin{bmatrix} 0 & \sigma^a \\ \bar{\sigma}^a & 0 \end{bmatrix}, \tag{8}$$

which satisfy the Clifford (or Dirac) algebra $\{\gamma^a, \gamma^b\} = 2\eta^{ab} I_{4\times 4}$, with $\eta^{ab}$ being the components of the inverse metric tensor in the Minkowski space-time $\eta^{ab} = \eta_{ab}^{-1} = \text{diag}(-1,1,1,1)$. However, for an arbitrary space-time with inverse metric tensor $g^{\mu\nu} = g_{\mu\nu}^{-1}$, the Clifford (or Dirac) algebra has to be rewritten into a general form $\{\gamma^\mu, \gamma^\nu\} = 2g^{\mu\nu} I_{4\times 4}$, where $\gamma^\mu = e_a^\mu \gamma^a$ is the Dirac matrix defined for general coordinate, with $e_a^\mu$ being the coframe field (correspondingly there is also a frame field $e_\mu^a$) describing the geometry of the local space-time. The frame $e_\mu^a$ and coframe $e_a^\mu$ fields have the orthogonal relations $e_a^\mu e_\mu^b = \delta_a^b$ and $e_a^\mu e_\nu^a = \delta_\nu^\mu$. The repeated indices are summed over according to the Einstein's summation rule [2] throughout this article. One can easily demonstrate the relation between the metric tensor and the frame field $g_{\mu\nu} = e_\mu^a e_\nu^b \eta_{ab}$ (and $g^{\mu\nu} = e_a^\mu e_b^\nu \eta^{ab}$) [2]. Here we note that the Greek indices ($\mu$, $\nu$, $\rho$, ...) represent the local coordinate indices *t, x, y* and *z*, and the Latin indices (*a, b, c,* …) represent the Lorentz frame indices 0, 1, 2 and 3. It is also notable that the coframe fields are vectors ($\partial_\mu$), while the frame fields are one-forms ($dx^\mu$), and as a result, the metric tensor is a two-form ($dx^\mu dx^\nu$), and its inverse is a two-vector ($\partial_\mu \partial_\nu$) [44]. The definition of the coframe field allows the Dirac equation to be rewritten into a very compact form $\gamma^\mu \partial_\mu \phi = 0$ [for chiral Weyl spinors we just need to replace $\gamma^\mu$ with $\bar{\sigma}^\mu$ ($\bar{\sigma}^\mu = e_a^\mu \bar{\sigma}^a$ positive) or $\sigma^\mu$ ($\sigma^\mu = e_a^\mu \sigma^a$ negative)], and hence Eq. (4) for the pair of Weyl points (*Q1* and *Q1'*) in our system defines a synthetic space-time geometry

$$e_0^\mu = (1, -w, 0, 0),\ e_1^\mu = (0, v_x, 0, 0),\ e_2^\mu = (0, 0, v_y, 0),\ e_3^\mu = (0, 0, 0, v_z). \tag{9}$$

If the components of coframe field are independent of the space coordinate (i.e. the homogeneous case), then we obtain a flat space-time for the pseudo-spinor. Oppositely, if the components are inhomogenously modulated, the synthetic space-time is curved, serving as the physical origin of synthetic gravitational field. For the latter case, the Weyl and Dirac equations (Eqs. 5-7) have to be reformulated, which will be introduced in detail in the following.

### III. Covariant form of Weyl and Dirac equations.

The eigen values of the Weyl equations [Eqs. (5) and (6)] can be easily solved as

$$\omega = \begin{cases} w\kappa_x + \sqrt{v_x^2 \kappa_x^2 + v_y^2 \kappa_y^2 + v_z^2 \kappa_z^2} \\ w\kappa_x - \sqrt{v_x^2 \kappa_x^2 + v_y^2 \kappa_y^2 + v_z^2 \kappa_z^2} \end{cases}, \tag{10}$$

where the partial differential operators $\partial_\alpha$ have been replaced with $i\kappa_\alpha$ ($\alpha = t, x, y, z$, $\kappa_t$ is actually the eigen frequency $\omega$), and thus $\kappa_\alpha$ ($\alpha = x, y, z$) is the wave vector relative to the momentum of Weyl degeneracy. By defining $v_{gx,\pm}$ as the group velocities of the transversal mode and the longitudinal mode near the Weyl cone in the $x$ direction, one can easily derive $v_{gx,+} = w+v_x$ and $v_{gx,-} = w-v_x$, respectively. Now we introduce the inhomogeneously controlled profile as displayed in FIG. 2(e). We assume $\omega_0 = p(k_x^2+k_y^2)+q$ to retain the time reversal symmetry of the system, with $p$ and $q$ being functions of the spatial coordinate. The inhomogeneity results in the variation of two parameters, $w$ and $v_x$, and hence we can elaborately design the inhomogeneity in space [$p(z)$ and $q(z)$], so that both $w$ and $v_x$ are linear functions along $z$ axis but independent of $x$ and $y$. Note that the parameters $p$ and $q$ are carefully tuned so that the positions and frequency of the Weyl points stay unchanged everywhere. Also the variation of $p$ and $q$ will not change $v_{gx,+}$, but will result in the variation of $v_{gx,-}$ in space. In this case, the space-time geometry for the pseudo-spinor is no longer flat [2] (i.e. the coframe field is spatially varying), and we need to reformulate the Weyl and Dirac equations so as to get a physical insight of the transport behaviors of the quasi-particles near the Weyl degeneracy.

### A. General coordinate and local Lorentz transformations of pseudo-spinors.

We first discuss the general coordinate transformation [e.g. a coordinate rotation of $\theta$ along $z$ axis], which will result in the variation of the constitutive parameters

$$\varepsilon(\theta) = \begin{bmatrix} 1+\frac{f_1}{\omega_0^2-\omega^2} & 0 & 0 \\ 0 & 1+\frac{f_1}{\omega_0^2-\omega^2} & 0 \\ 0 & 0 & 1 \end{bmatrix}, \mu(\theta) = \begin{bmatrix} 1+\frac{f_2\omega^2}{\omega_0^2-\omega^2} & 0 & 0 \\ 0 & 1+\frac{f_2\omega^2}{\omega_0^2-\omega^2} & 0 \\ 0 & 0 & 1 \end{bmatrix}, \gamma(\theta) = \begin{bmatrix} -\frac{\sqrt{f_1 f_2}\omega\cos 2\theta}{\omega_0^2-\omega^2} & \frac{\sqrt{f_1 f_2}\omega\sin 2\theta}{\omega_0^2-\omega^2} & 0 \\ \frac{\sqrt{f_1 f_2}\omega\sin 2\theta}{\omega_0^2-\omega^2} & \frac{\sqrt{f_1 f_2}\omega\cos 2\theta}{\omega_0^2-\omega^2} & 0 \\ 0 & 0 & 0 \end{bmatrix}. \quad (11)$$

Hence, there will be a $-\theta$ rotation of the full band structure, as well as the Weyl cones with respect to the new coordinate system ($x'y'$). For the pair of Weyl points $Q1$ and $Q1'$, a representation of coordinate rotation can be obtained

$$e'^\nu_a = \frac{\partial x'^\nu}{\partial x^\mu} e^\mu_a, \quad \gamma'^\nu = \frac{\partial x'^\nu}{\partial x^\mu} \gamma^\mu, \tag{12}$$

where the transformation matrix is expressed in the form

$$\frac{\partial x'^\nu}{\partial x^\mu} = \begin{bmatrix} 1 & 0 & 0 & 0 \\ 0 & \cos\theta & -\sin\theta & 0 \\ 0 & \sin\theta & \cos\theta & 0 \\ 0 & 0 & 0 & 1 \end{bmatrix}. \tag{13}$$

Equations (12-13) provide a transformation rule for the general coordinate indices $\mu, \nu = t, x, y, z$, and the form of Eq. (13) can be other 4 by 4 matrix forms denoting SO(3) rotations or Lorentz boosts (the general form is Eq. D12 by replacing $ab$ with $\mu\nu$), which depends on the transformation operation on the full Hamiltonian.

Next, we discuss the local Lorentz transformation for the pseudo-spinor. Different from the general coordinate transformation, the local Lorentz transformation will not change the coordinate system, but will result in the variation of local Lorentz indices ($a, b = 0,1,2,3$). The Hamiltonian can be divided into several blocks as shown in Eq. (4), and thus the local transformation operator [e.g. an SO(2) transformation of $x$ and $y$ components of field vectors] is a block diagonal form

$$U = \begin{bmatrix} U_2 & 0 & 0 & 0 \\ 0 & U_2 & 0 & 0 \\ 0 & 0 & U_1 & 0 \\ 0 & 0 & 0 & U_1 \end{bmatrix}. \tag{14}$$

with $U_1$ and $U_2$ being a 2 by 2 and a 3 by 3 matrices characterizing SO(2) transformation respectively (thus $UU^\dagger=1$), acting as the transformation operation to each block in the full Hamiltonian and to the vector fields of eigen state (Appendix A). It is notable that $U$ acts on $\psi$ instead of $\tilde{\psi}$, and hence by applying the transformation, we obtain

$UMU^\dagger \psi'(x^\mu) = \omega UNU^\dagger \psi'(x^\mu)$, where $\psi'$ is the transformed eigenstate $\psi' = U\psi$. For local transformations, we do not need to apply the transformation on the coordinate $x^\mu$. This is essentially different from the general coordinate transformation in previous discussions. It can be easily demonstrated that the commutation $[U, N] = 0$ is satisfied, and thus the Hamiltonian and eigen state transform as

$$H' = N^{-1/2}UMU^\dagger N^{-1/2} = THT^\dagger, \quad \tilde{\psi}' = N^{1/2}U\psi = T\tilde{\psi}. \tag{15}$$

with the transformation operator $T = N^{1/2}UN^{-1/2}$. By applying the k.p perturbation theory, the effective Hamiltonian also obey a transformation rule (Appendix B),

$$H'_{\text{eff}} = LH_{\text{eff}}L^\dagger, \tag{16}$$

where $L$ is the effective local transformation operator with elements $(L)_{ij} = \langle \tilde{\psi}_i | T | \tilde{\psi}_j \rangle$ (as well as its complex conjugate $(L^\dagger)_{ij} = \langle \tilde{\psi}_j | T^\dagger | \tilde{\psi}_i \rangle$), and is also unitary. Considering the two bands forming the Weyl degeneracy in the positive frequency domain, the Lie algebraic infinitesimal of $L$ ($\delta L$ being a 2 by 2 matrix obtained from Lie algebra $\delta L = dL/d\lambda|_{\lambda=0}$, with $\lambda$ being a coefficient in transformation operator) is anti-Hermitian with basis $i\sigma^a$ ($a$=1,2,3). If the operator $U$ characterizes Lorentz boost, $U$ is not a unitary operator but still has determinant 1, via similar processes, one obtains a Hermitian infinitesimal $\delta L$ with basis $\sigma^a$ ($a$=1,2,3) (Appendix B), and thus a general form of the basis can be provided $\sigma^{ab} = 1/2(\sigma^a\bar{\sigma}^b - \sigma^b\bar{\sigma}^a)$. For the two bands in the negative frequency domain, the basis is simply $\bar{\sigma}^{ab} = 1/2(\bar{\sigma}^a\sigma^b - \bar{\sigma}^b\sigma^a)$. The discussions allow a definition of the local Lorentz transformation of the Dirac pseudo-spinor

$$L(\lambda) = \exp(\frac{1}{4}\lambda_{ab}\gamma^{ab}), \tag{17}$$

with $\gamma^{ab}$ being the basis defined on symmetry group SL(2,C)

$$\gamma^{ab} = \frac{1}{2}\begin{bmatrix} \sigma^a\bar{\sigma}^b - \sigma^b\bar{\sigma}^a & 0 \\ 0 & \bar{\sigma}^a\sigma^b - \bar{\sigma}^b\sigma^a \end{bmatrix} = \frac{1}{2}[\gamma^a, \gamma^b]. \tag{18}$$

Here both $\lambda_{ab}$ and $\gamma^{ab}$ are antisymmetric about the indices $a$ and $b$. It is noted that the transformation is definitely the representation of Lorentzian group for Dirac spinors [2], and hence the Lorentz transformation of Dirac pseudo-spinors is well defined. For Weyl pseudo-spinors, since Eq. (18) is block diagonalized, we just need to consider the blocks as the basis of local Lorentz transformation. The local transformation can be used to derive Landau quantization in tilted Weyl cone systems with uniform magnetic field background [28].

### B. Covariant derivatives.

Since the transformation rules are well defined as discussed in Subsection III. A, we are now ready to investigate the covariant forms in detail. Here we need to borrow the concept of parallel transport in differential geometry [44]. With the introduction of inhomogeneous modulations (FIG. 2e), the coframe field (Eq. 9) is spatially varying, and hence the metric tensor $g^{\mu\nu}$ for the synthetic space-time is also a function of spatial coordinate. According to the differential geometry, every Lie group defines a manifold [44], and hence the transformation rule allows the definition of a curved Lorentz manifold as a result of inhomogeneous modulations. This is actually a synthetic curved space-time geometry for the pseudo-spinor, with the transformation operator depends on space [i.e. operator $L(\lambda)$ is spatially dependent $\lambda(x^\mu)$]. If now we apply a local transformation $L(\lambda)$ to the pseudo-spinor, the action of partial derivative will generate two terms [i.e. $\partial_\mu(L\varphi) = (\partial_\mu L)\varphi + L(\partial_\mu\varphi)$]. Hence $\partial_\mu\varphi$ no longer transforms as a Dirac pseudo-spinor, and the transformation rule is changed. To maintain the transformation rule, the partial differential operator has to be replaced with a covariant derivative $\partial_\mu \to \nabla_\mu$, where the operator $\nabla_\mu$ for Dirac pseudo-spinor is expressed as

$$\nabla_\mu \phi = \partial_\mu \phi + \frac{1}{4}\omega_{\mu ab}(z)\gamma^{ab}\phi. \tag{19}$$

The second term is a non-Abelian gauge field defined on symmetry group SL(2,C), with the $\omega_{\mu ab}$ being the spin connection that is antisymmetric about the indices $a$ and $b$ [2]. The spin connection is a function of $z$ because the designed system is inhomogeneous in $z$ direction. For Weyl pseudo-spinors, the covariant derivatives are simply $\nabla_\mu = \partial_\mu + 1/2\,\omega_{\mu ab}(z)\sigma^{ab}$ and $\nabla_\mu = \partial_\mu + 1/2\,\omega_{\mu ab}(z)\bar\sigma^{ab}$ for partners with oppsite chiralities respectively, owing to the fact that Eq. (19) is block diagonalized. The spin connection (as well as the non-Abelian gauge field) is also a one-form, which is connected with the frame field as

$$T^a = de^a + \omega^a{}_b \wedge e^b. \tag{20}$$

Here $e^a$ and $\omega^a{}_b$ are the one-forms of the frame field $e^a = e^a_\mu dx^\mu$ and spin connection $\omega^a{}_b = \omega^a_{\mu b} dx^\mu$, $d$ denotes total differential, $T^a$ is the torsion two-form $T^a = T^a_{\mu\nu} dx^\mu x^\nu$, and $\wedge$ is the wedge product. The spin connection can be solved from Eq. (20) by taking $T^a$ and $e^a$ as knowns, and the details can be found in Appendix B. It is notable that the indices can be raised or lowered by the metric tensor or the frame field (e.g. $\omega^a_{\mu\ b} = \omega_{\mu cb}\eta^{ca}$, $\omega^{\ \rho}_{\mu\ \sigma} = \omega_{\mu\tau\sigma}g^{\tau\rho}$, $\omega^{\ \rho}_{\mu\ \sigma} = \omega_{\mu ab}e^{a\rho}e^b_\sigma$ etc). One can easily identify that by applying a local transformation, the following relation can be satisfied

$$\nabla_\mu \phi \to (\nabla_\mu \phi)' = \nabla'_\mu(L\phi) = L\nabla_\mu \phi, \tag{21}$$

where $\nabla'_\mu$ is the transformed covariant derivative $\nabla'_\mu = L\nabla_\mu L^{-1}$ acting on Dirac pseudo-spinors, including a spin connection term with the transformed form

$$\omega'^{a}{}_{b} = \Lambda^{-1a}{}_{b} d\Lambda^{c}{}_{d} + \Lambda^{-1a}{}_{c} \omega^{c}{}_{d} \Lambda^{d}{}_{b}. \tag{22}$$

Here $\Lambda^{a}{}_{b}$ is the component of the 4 by 4 transformation matrix acting on Lorentzian vectors (with indices $a$=0,1,2,3), and we put the detailed discussion of $\Lambda^{a}{}_{b}$ into Appendix D to avoid redundancy. Equation (21) is the basic requirement of parallel transport of a spinor on a curved manifold [44]. As a consequence, Eq. (22) is the general transformation rule of non-Abelian gauge fields [1,2].

**IV. Analysing chiral transport behaviour by solving covariant form of Weyl equations.**

In the last Section, we discussed in detail how to construct the Weyl and Dirac equations for inhomogeneous systems that the group velocity (or frame field) near the Weyl cone is spatially varying. To get a physical insight on the transport behavior of the pseudo-spinors, in this section we will try to solve this covariant form of equations, the solutions of which enables us to analyze the propagation direction of the bulk modes supported by the inhomogeneous system. Since the covariant form of Dirac equation is block diagonalized, and also researchers are mainly concerned with the Weyl cone in the positive frequency domain, here we only present the result for covariant form of Weyl equation $\bar{\sigma}^\mu \nabla_\mu \varphi = 0$.

We first calculate the components of non-Abelian gauge field (i.e. spin connection) in Eq. (19). The torsion two-form defined in Eq. (20) is a new degree of freedom that characterizes the spin of matter source that emits gravitational field. Our designed system is a very simple case that the group velocity is only varying in one direction, which corresponds a spinless source of matter. Hence, the spin connection is torsion free. It should be noted that the spin connection cannot be zero (even though it is not obvious in view of *k.p* perturbation theory), and if the spin connection is ignored, there will be serious consequences. In addition, the parallel transport cannot be satisfied if the spin connection is zero. However, this term is often ignored in some previous works [19,32,33], and we will discuss this in detail in Appendix E. Solving Eq. (20) (or Eq. C4), one can obtain the general expression of spin connection (independent of torsion or torsion free)

$$\omega_{\mu\ b}^{\ a} = e^a_\nu \partial_\mu e^\nu_b + e^a_\nu \Gamma^\nu_{\mu\sigma} e^\sigma_b, \tag{23}$$

where $\Gamma^\nu_{\mu\sigma}$ is the affine connection. We previously understood that the form of the connection (or gauge field) in the covariant derivative is determined by the transformation rule of the spinor. Hence, for other formed spinors, the connections (or gauge field) will be different, such as a general coordinate contravariant vector field $\nabla_\mu V^\nu = \partial_\mu V^\nu - \Gamma^\nu_{\mu\rho} V^\rho$ and a Lorentz covariant vector field $\nabla_\mu V_a = \partial_\mu V_a + \omega_{\mu ab} V^b$. This is because $V^\nu$ is characterized by the transformation rule in Eqs. (12-13), but $V_a$ transforms as $V_a \to V'_a = \Lambda^a_{\ b} V^b$ ($\Lambda^a_{\ b}$ has been defined following Eq. 22). For the torsion free case, $\Gamma^\nu_{\mu\sigma}$ reduces to the Levi Civita connection, and is expressed as

$$\Gamma^\rho_{\mu\nu}(g) = \frac{1}{2} g^{\rho\sigma} (\partial_\mu g_{\sigma\nu} + \partial_\nu g_{\mu\sigma} - \partial_\sigma g_{\mu\nu}). \tag{24}$$

The metric tensor and its inverse have been given in Appendix F. It is found from Eq. (24) that the Levi Civita connection is symmetric about the subscript indices ($\mu,\nu$). For our designed system (taking Weyl point $Q1$ as an example, Eq. 9), the following components are non-zero

$$\begin{aligned}
&\Gamma^t_{tz} = \Gamma^t_{zt} = -\frac{1}{2} w' w v_x^{-2}, \Gamma^t_{xz} = \Gamma^t_{zx} = -\frac{1}{2} w' v_x^{-2}, \\
&\Gamma^x_{tz} = \Gamma^x_{zt} = \frac{1}{2} w' + \frac{1}{2} w' w^2 v_x^{-2} - v'_x w v_x^{-1}, \Gamma^x_{xz} = \Gamma^x_{zx} = \frac{1}{2} w' w v_x^{-2} - v'_x v_x^{-1}, \\
&\Gamma^z_{tt} = -w' w v_x^{-2} v_z^2 + v'_x w^2 v_x^{-3} v_z^2, \Gamma^z_{tx} = \Gamma^z_{xt} = -\frac{1}{2} w' v_x^{-2} v_z^2 + v'_x w v_x^{-3} v_z^2, \Gamma^z_{xx} = v'_x v_x^{-3} v_z^2
\end{aligned} \tag{25}$$

Substituting Eq. (25) into Eq. (24), we derive the nonzero components of spin connection

$$\omega_{t[03]} = \frac{1}{2}w'wv_x^{-2}v_z, \quad \omega_{t[13]} = \frac{1}{2}w'v_x^{-1}v_z - v_x'wv_x^{-2}v_z,$$
$$\omega_{x[03]} = \frac{1}{2}w'v_x^{-2}v_z, \quad \omega_{x[13]} = -v_x'v_x^{-2}v_z, \quad \omega_{z[01]} = \frac{1}{2}w'v_x^{-1}$$
(26)

where the bracket […] denotes the anti-symmetric pair of indices, and the prime denotes the first order partial derivative with respect to $z$. These terms can be used to solve the covariant form of Weyl equation for our system. Similar to the magnetic field, the commutation of two covariant derivatives defines the space-time curvature for the pseudo-spinor $[\nabla_\mu, \nabla_\nu]\varphi = 1/4 R_{\mu\nu ab}\sigma^{ab}\varphi$, where $R_{\mu\nu ab}$ is the component of Riemannian curvature two-form $R_{\mu\nu}dx^\mu dx^\nu$, and can be obtained with the Yang-Mills gauge theory

$$R_{\mu\nu ab} = \partial_\mu \omega_{\nu ab} - \partial_\nu \omega_{\mu ab} + \omega_{\mu ac}\omega_{\nu\ b}^{\ c} - \omega_{\nu ac}\omega_{\mu\ b}^{\ c}.$$
(27)

We note that the Riemannian curvature is anti-symmetric about the indices $\mu$ and $\nu$, and thus has 6 components. Each component is a 4 by 4 matrix with components denoted by indices $a$ and $b$ (antisymmetric about $a$ and $b$), and can be obtained as the following expressions for Weyl point *Q1*

$$R_{tx[01]} = \frac{1}{4}w'^2v_x^{-3}v_z^2, \quad R_{tz[03]} = -\frac{3}{4}w'^2v_x^{-2}v_z + \frac{3}{2}w'v_x'wv_x^{-3}v_z,$$
$$R_{tz[13]} = \frac{3}{2}w'v_x'v_x^{-2}v_z - 2v_x'^2wv_x^{-3}v_z - \frac{1}{4}w'^2wv_x^{-3}v_z,$$
(28)
$$R_{xz[03]} = \frac{3}{2}w'v_x'v_x^{-3}v_z, \quad R_{xz[13]} = -2v_x'^2v_x^{-3}v_z - \frac{1}{4}w'^2v_x^{-3}v_z,$$

The Riemannian curvature plays a similar role as the magnetic field [the curvature two-form on a curved U(1) manifold], which leads to the quantization of energy levels of Weyl spinors. For the other Weyl points (*Q2-Q4*), we have provided the frame field in Appendix F, and the analysing processes are similar to *Q1*, which will not be repeated here.

Different from a Weyl pseudo-spinor in a background magnetic field (see the comparison in Appendix D), the covariant form of Weyl equation coupling with gravitational field cannot be solved analytically. We instead use the PDE module of COMSOL (see Methods) to numerically solve the quantized energy levels, with results shown in FIG. 3. Here $\partial_x$ and $\partial_y$ are replaced with $i\kappa_x$ and $i\kappa_y$ because they are both good quantum numbers. The insets in the figure clearly show the quantization of energy levels induced by the curvature. In contrast to the Landau quantization resulting from an axial magnetic field, the energy levels are dispersive in both $\kappa_x$ and $\kappa_y$ directions. The zeroth energy levels of different Weyl points propagate in different chiral axial directions - *Q1* (*Q2*) propagates in positive *y* (*x*) direction, and *Q3* (*Q4*) propagates in negative *y* (*x*) direction, which leads to gravitational anomaly [30,31,39]. On different $\kappa_x$ (for *Q1* and *Q3*) or $\kappa_y$ (for *Q2* and *Q4*) planes, the quantized energy levels of *Q1* (*Q2*) are shifted towards higher frequency as $\kappa_x$ ($\kappa_y$) varies from negative to positive (see FIG. 3a1-a5) [conversely for *Q3* (*Q4*), see FIG. 3b1-b5]. The dispersion of zeroth chiral mode becomes nonlinear in the axial direction ($\kappa_y$ for *Q1* and *Q3*, $\kappa_x$ for *Q2* and *Q4*, see panels 1-2 and 4-5) out of $\kappa_x(\kappa_y)$ plane, meanwhile the group velocity of the zeroth order mode in *y* (*x*) direction for *Q1* and *Q3* (*Q2* and *Q4*) decreases gradually to zero [at $\kappa_y(\kappa_x)=0$] as $\kappa_x$ ($\kappa_y$) deviates from zero. The frequency difference between adjacent energy levels tends to increase as $\kappa_x$ ($\kappa_y$) deviates from zero at high values of $\kappa_y$ ($\kappa_x$) planes for *Q1* and *Q3* (*Q2* and *Q4*) (see panels 1-2 and 4-5 in FIG. 3). From the analysis, one can easily find the transport properties of the zeroth mode as displayed in FIG. 4, where $v_{g1} \sim v_{g4}$ represent group velocities of zeroth mode for *Q1* to *Q4*, respectively. The insets in FIG. 4 show the 2D dispersions of zeroth modes, with the colour scales showing the equi-frequency contour. Both FIG. 3 and FIG. 4 show that the group velocities of zeroth modes have the following relations, $v_{g1(2)} = -v_{g3(4)}$ and $v_{g1,y(x)} = v_{g2,x(y)}$, with the subindices $x$ and $y$ denoting the components. Since the two mirror planes (*M1* and *M2* as enclosed by dashed lines in FIG. 4) in the system remains intact after the introduction of the inhomogeneous modulation in the $z$ direction, the relations of group velocities also obey the mirror symmetry of the system.

## V. Conceptual sample design

The unit cell structure in FIG. 5(a) can realize the constitutive parameters in Eq. (1) at the long wavelength limit [22]. The unit cell is composed of two dielectric layers with dielectric constants $\varepsilon_1$ and $\varepsilon_2$, and with thicknesses $h_1$ and $h_2$ respectively. Two subwavelength metallic split rings, which can be treated as perfect electric conductors (PEC) at microwave frequencies, are aligned orthogonally and placed inside the first dielectric layer. The structure has two mirror symmetric planes in the diagonal directions with respect to $x$ and $y$ axis (as enclosed by *M1* and *M2* dashed lines in the figure). This is consistent with the symmetry of the constitutive parameters of the metamaterial. Figure 5(b) shows the Brillouin zone of the metamaterial. The band structures in the Γ-M direction are calculated numerically, and shown in FIG. 5(c)-(e) (impact of loss is discussed in Appendix G). We find that the Weyl point arises from the crossing between the second and the third energy bands, agreeing well with prediction from the constitutive parameters (see FIG. 2 and FIG. 6). The nearly flat band corresponds to the longitudinal mode of the homogenized metamaterial. The longitudinal mode becomes dispersive due to the coupling of the subwavelength resonators, and such a dispersion cannot be described using local effective medium theory, but can be handled using non-local effective media theory in which the constitutive parameters becomes **k**-dependent [43]. Hence, tuning the dispersion of the nearly flat band is equivalent to the modulation of nonlocality of constitutive parameters (Eq. 1). The resonance frequencies at Γ and M points are different from each other, which can be tuned by adjusting the coupling strength of metallic resonators between adjacent unit cells in $z$ direction. By gradually increasing the thickness $h_2$, the transition of Weyl cone from Type-I to Type-II is achieved, as shown in FIG. 5(c)-(e). The thickness $h_2$ can be inhomogeneously tuned along the $z$ direction to generate the synthetic gravitational field for Weyl pseudo-spinors.

## VI. Summary

In summary, we have proposed that an inhomogeneous Weyl metamaterial with spatially controlled non-locality can manifest chiral gravitational effects. Such a system has a spatially dependent longitudinal band dispersion, which generates a synthetic curved space-time geometry (or gravitational field) background for the Weyl pseudo-spinors. We showed that Riemannian curvature, derived from a non-Abelian gauge field associated with spin connection, can give rise to the quantization of the energy levels of pseudo-spinors. The zeroth order energy level is propagative in the chiral directions, and is responsible for the chiral gravitational effect in this inhomogeneous system. Our theory offers the first realistic scheme for implementing synthetic gravitational field and for observing chiral gravitational effect for Weyl pseudo-spinors in photonic systems. Such a system can be realized in table-top experiments, and it should be an ideal platform for experimentally verifying the theories that describes the interaction between particles and gravitational field.

## Acknowledgements

This work is supported by Hong Kong RGC (AoE/P-02/12, 16304717). S. Zhang acknowledges the financial support from the European Research council Consolidator Grant (Topological) and Horizon 2020 Action Project grant 734578 (D-SPA). We acknowledge Prof. C. X. Liu for helpful suggestions in the model construction.

**Appendix A: Full and effective Hamiltonians**

The Maxwell equation for a homogenous system can be written as an eigen value problem [42], and in the following we show the details for deriving the Hermitian Hamiltonian. The constitutive relation given by Eq. (2) can be expanded as

$$
\begin{aligned}
P_x &= \frac{\varepsilon_0 f_1}{\omega_0^2 - \omega^2} E_x - \frac{i\sqrt{f_1 f_2}\,\omega}{c(\omega_0^2 - \omega^2)} H_x, \quad P_y = \frac{\varepsilon_0 f_1}{\omega_0^2 - \omega^2} E_y + \frac{i\sqrt{f_1 f_2}\,\omega}{c(\omega_0^2 - \omega^2)} H_y \\
M_x &= \frac{\mu_0 f_2 \omega^2}{\omega_0^2 - \omega^2} H_x + \frac{i\sqrt{f_1 f_2}\,\omega}{c(\omega_0^2 - \omega^2)} E_x, \quad M_y = \frac{\mu_0 f_2 \omega^2}{\omega_0^2 - \omega^2} H_y - \frac{i\sqrt{f_1 f_2}\,\omega}{c(\omega_0^2 - \omega^2)} E_y
\end{aligned}
\tag{A1}
$$

where $c$ is the velocity of light, with $c = 1/\sqrt{\varepsilon_0 \mu_0}$. The source-free Maxwell equation in the frequency domain can be recasted into the following form

$$-ic\nabla \times \sqrt{\varepsilon_0}\,\mathbf{E} = \omega\sqrt{\mu_0}\,\mathbf{H} + \omega\sqrt{1/\mu_0}\,\mathbf{M}, \quad ic\nabla \times \sqrt{\mu_0}\,\mathbf{H} - \omega\sqrt{1/\varepsilon_0}\,\mathbf{P} = \omega\sqrt{\varepsilon_0}\,\mathbf{E}, \tag{A2}$$

The relation between $\mathbf{P}$ and $\mathbf{M}$ can be derived from Eq. (A1)

$$M_x = i\frac{\sqrt{\mu_0 f_2}\,\omega}{\sqrt{\varepsilon_0 f_1}} P_x, \quad M_y = -i\frac{\sqrt{\mu_0 f_2}\,\omega}{\sqrt{\varepsilon_0 f_1}} P_y, \tag{A3}$$

$$
\begin{aligned}
-i\sqrt{\varepsilon_0 \mu_0}\sqrt{f_1 f_2}\,E_x + i\omega_0^2 \frac{\sqrt{\mu_0 f_2}}{\sqrt{\varepsilon_0 f_1}} P_x &= \omega M_x + \mu_0 f_2 \omega H_x \\
-i\omega_0^2 \frac{\sqrt{\mu_0 f_2}\,\omega}{\sqrt{\varepsilon_0 f_1}} P_y + i\sqrt{\varepsilon_0 \mu_0}\sqrt{f_1 f_2}\,E_y &= \omega M_y + \mu_0 f_2 \omega H_y
\end{aligned}
\tag{A4}
$$

The linear equations [Eqs. (A2-A4)] can be written into the matrix form $M\psi = \omega N\psi$ as given by Eq. (3) in the main text, where the blocks in $M$ and $N$ are

$$K = \begin{bmatrix} 0 & -ic\partial_z & ic\partial_y \\ ic\partial_z & 0 & -ic\partial_x \\ -ic\partial_y & ic\partial_x & 0 \end{bmatrix}, T_1 = -\sqrt{f_1/f_2}\begin{bmatrix} 1 & 0 \\ 0 & -1 \\ 0 & 0 \end{bmatrix}, T_2 = -\omega_0^2/\sqrt{f_1 f_2}\sigma^3, T_1' = \begin{bmatrix} 1 & 0 \\ 0 & 1 \\ 0 & 0 \end{bmatrix}, \quad (A5)$$

If we apply a transformation to the eigen state $\psi \to \psi' = U\psi$ [$U$ is defined in Eq. (10), with blocks acting on vector fields $U_2\mathbf{E}$, $U_2\mathbf{H}$, $U_1\mathbf{P}$ and $U_1\mathbf{M}$], the blocks in $M$ transforms as $K' = U_2 K U_2^\dagger$, $T_1' = U_2 T_1 U_1^\dagger$. Hence, solving the plane wave solutions is casted into an eigen value problem $H\tilde{\psi} = \omega\tilde{\psi}$, with the exact forms expressed in Eq. (3). The dispersion diagram by solving the eigen values of the Hamiltonian is shown in FIG. 6. We find that each Weyl point ($Q1$-$Q4$) has its partner with opposite chirality located in the negative frequency domain, with the same $\mathbf{k}$ vector in momentum space.

The effective Hamiltonian describing the Weyl cone can be obtained with k.p theory. In homogeneous systems, the momentum in three directions are good quantum numbers, and thus the partial derivatives $\partial_x$, $\partial_y$ and $\partial_z$ can be replaced by $ik_x$, $ik_y$ and $ik_z$, respectively. We first expand the full Hamiltonian near the Weyl point

$$H(\mathbf{k}) = H(\mathbf{k}^W + \boldsymbol{\kappa}) = H(\mathbf{k}^W) + \sum_{\alpha=x,y,z} \frac{\partial H}{\partial \kappa_\alpha} \kappa_\alpha, \quad (A6)$$

where $\mathbf{k}^W = (\pm k_x^W, 0, 0)$ or $(0, \pm k_y^W, 0)$ is the $\mathbf{k}$ of Weyl points, and $\boldsymbol{\kappa} = (\kappa_x, \kappa_y, \kappa_z)$ is the $\mathbf{k}$ relative to Weyl points momentum $\boldsymbol{\kappa} = \mathbf{k} - \mathbf{k}^W$. If the eigenstate at $\mathbf{k}^W$ is denoted as $\tilde{\psi}_i^W$, then the eigenstate $\tilde{\psi}_i$ at $\mathbf{k}$ can be expanded $\tilde{\psi}_i = \sum_{j=1}^J C_{i,j} \tilde{\psi}_j^W$, with $J$ being the dimension of full Hamiltonian. The coefficient $C_{i,j}$ is defined by the inner product $C_{i,j} = \langle \tilde{\psi}_j^W | \tilde{\psi}_i \rangle$. The two eigenstates forming the

Weyl degeneracies in the positive frequency region are assigned with sub-indices $i = 1, 2$, satisfying $H(\mathbf{k}^W)\tilde{\psi}_i^W = \omega \tilde{\psi}_i^W$. Inserting the identity operator $I = \sum_j |\tilde{\psi}_j^W\rangle\langle\tilde{\psi}_j^W|$, the components of the effective Hamiltonian can thus be obtained as

$$(H_{\text{eff}})_{ij} = \langle\tilde{\psi}_i^W|H|\tilde{\psi}_j^W\rangle = \omega_W \delta_{ij} + \sum_{\alpha=x,y,z}\langle\tilde{\psi}_i^W|\partial_{\kappa_\alpha}H|\tilde{\psi}_j^W\rangle \kappa_\alpha. \tag{A7}$$

The first term denotes the Weyl frequency, and since it is just a constant, we will neglect it in the remaining of the text. The second term can be decomposed as in the following. We set the matrix $P_\alpha$ having the components $P_{\alpha,ij} = \langle\tilde{\psi}_i^W|\partial_{\kappa_\alpha}H|\tilde{\psi}_j^W\rangle$. Since $P_\alpha$ is Hermitian, it can be written as the sum of Pauli matrices,

$$P_\alpha = w_\alpha \kappa_\alpha \bar{\sigma}^0 + \sum_{b=1,2,3} v_{\alpha b} \kappa_\alpha \bar{\sigma}^b, \tag{A8}$$

with the coefficients derived as $w_\alpha = -1/2 Tr(P_\alpha)$ and $v_{\alpha b} = 1/2 Tr(P_\alpha \sigma^b)$. These coefficients can be numerically calculated by the diagonalization of the Hamiltonian in Eq. (3), and the result shows that $w_\alpha$ have only one nonzero component $w_x$, and $v_{\alpha b}$ have three nonzero components $v_{x1}$, $v_{y2}$ and $v_{z3}$. Taking Weyl point $Q1$ as an example, $v_{x1}$, $v_{y2}$ and $v_{z3}$ are positive and $w_x$ is negative. We set $w = -w_x$, $v_x = v_{x1}$, $v_y = v_{y2}$ and $v_z = v_{z3}$ (meaning $w$, $v_x$, $v_y$ and $v_z$ are all positive defined), and thus the effective Hamiltonian describing the Weyl cone $Q1$ can be derived as

$$H_{Q1} = -w\bar{\sigma}^0 \kappa_x + v_x \bar{\sigma}^1 \kappa_x + v_y \bar{\sigma}^2 \kappa_y + v_z \bar{\sigma}^3 \kappa_z. \tag{A9}$$

By replacing $\kappa_\alpha$ with the operator $-i\partial_\alpha$, and $\omega$ replaced with $-i\partial_t$, we have Eq. (5), and Eq. (6) is obtained via similar processes. The effective Hamiltonian and the $C$ matrix with components $C_{i,j} = \langle\tilde{\psi}_j^W|\tilde{\psi}_i\rangle$ can be associated with an eigenvalue problem $H_{\text{eff}}C_i = \omega_i C_i$ ($C_i$ is a column in $C$). Hence, the two eigenstates of the Weyl equation ($\varphi$ in Eq. 5) are just the two column vectors

($i$=1,2) in $C$ matrix by picking the corresponding two components. The spinor with opposite chirality ($\chi$ in Eq. 6) is obtained in a similar way.

**Appendix B: k.p effective transformation operators**

It is well known that the effective Hamiltonian (Weyl Hamiltonian) can be obtained with *k.p* perturbation theory (Appendix A). For local transformations, the matrix form acting on the full Hamiltonian can be reduced to an effective form (2 by 2 matrix), which can also be obtained with the *k.p* perturbation theory. The transformed Hamiltonian $H'=THT^\dagger$ (e.g. $T$ is a matrix characterizing SO(2) transformation, now $T=N^{1/2}UN^{-1/2}$ is a unitary operator and $TT^\dagger=1$, and thus $\tilde{\omega}=\omega TT^\dagger=\omega$) can be expanded near the Weyl point

$$H'(\mathbf{k})=H'(\mathbf{k}^W+\mathbf{\kappa})=H'(\mathbf{k}^W)+\sum_{\alpha=x,y,z}\frac{\partial H'}{\partial \kappa_\alpha}\kappa_\alpha, \tag{B1}$$

and thus we have

$$H'(\mathbf{k}^W+\mathbf{\kappa})\left|\tilde{\psi}'_a(\mathbf{k}^W+\mathbf{\kappa})\right\rangle=H'(\mathbf{k}^W)\left|\tilde{\psi}'_a(\mathbf{k}^W+\mathbf{\kappa})\right\rangle+\sum_{\alpha=x,y,z}\frac{\partial H'}{\partial \kappa_\alpha}\left|\tilde{\psi}'_a(\mathbf{k}^W+\mathbf{\kappa})\right\rangle\kappa_\alpha. \tag{B2}$$

Applying the identity operator $I=\sum_a\left|\tilde{\psi}_a(\mathbf{k}^W)\right\rangle\left\langle\tilde{\psi}_a(\mathbf{k}^W)\right|=\sum_a\left|\tilde{\psi}'_a(\mathbf{k}^W)\right\rangle\left\langle\tilde{\psi}'_a(\mathbf{k}^W)\right|$, Eq. (B2) reads

$$\sum_b\left|\tilde{\psi}'_b(\mathbf{k}^W)\right\rangle\left\langle\tilde{\psi}'_b(\mathbf{k}^W)\right|\omega_a(\mathbf{k}^W+\mathbf{\kappa})\left|\tilde{\psi}'_a(\mathbf{k}^W+\mathbf{\kappa})\right\rangle$$
$$=\sum_b\left|\tilde{\psi}'_b(\mathbf{k}^W)\right\rangle\left\langle\tilde{\psi}'_b(\mathbf{k}^W)\right|\omega_b(\mathbf{k}^W)\left|\tilde{\psi}'_a(\mathbf{k}^W+\mathbf{\kappa})\right\rangle+\sum_{b,\alpha}\left|\tilde{\psi}'_b(\mathbf{k}^W)\right\rangle\left\langle\tilde{\psi}'_b(\mathbf{k}^W)\right|\frac{\partial H'}{\partial \kappa_\alpha}\left|\tilde{\psi}'_a(\mathbf{k}^W+\mathbf{\kappa})\right\rangle\kappa_\alpha. \tag{B3}$$

Now we are ready to insert the transformation operator [$\tilde{\psi}_b(\mathbf{k}^W)=T^\dagger\tilde{\psi}'_b(\mathbf{k}^W)$] into Eq. (B3), and simple derivations lead to

$$\omega_a(\mathbf{k}^W+\mathbf{\kappa})\left|\tilde{\psi}_b(\mathbf{k}^W)\right\rangle=\omega_b(\mathbf{k}^W)\left|\tilde{\psi}_b(\mathbf{k}^W)\right\rangle+\sum_\alpha\frac{\partial H'}{\partial \kappa_\alpha}\left|\tilde{\psi}_b(\mathbf{k}^W)\right\rangle\kappa_\alpha. \tag{B4}$$

Product of $\langle \tilde{\psi}_b(\mathbf{k}^W)|$ from the left gives

$$\omega_a(\mathbf{k}^W + \boldsymbol{\kappa}) = \omega_b(\mathbf{k}^W)\delta_{ab} + \sum_\alpha \langle \tilde{\psi}_a(\mathbf{k}^W)| \frac{\partial H'}{\partial \kappa_\alpha} |\tilde{\psi}_b(\mathbf{k}^W)\rangle \kappa_\alpha, \tag{B5}$$

and the transformed effective Hamiltonian can be obtained

$$(H'_{\text{eff}})_{ab} = \omega_W \delta_{ab} + \sum_{\alpha=x,y,z} \langle \tilde{\psi}_a | T(\partial_{\kappa_\alpha} H) T^\dagger | \tilde{\psi}_b \rangle \kappa_\alpha, \tag{B6}$$

The first term is a constant term and is out of consideration here, and by inserting the identity $I = \sum_a |\tilde{\psi}_a(\mathbf{k}^W)\rangle\langle \tilde{\psi}_a(\mathbf{k}^W)|$, the transformed effective Hamiltonian is derived as Eq. (16). The discussions above considers an SO(2) rotation as an example, and thus the transformation operator $T$ and the effective operator $L$ are both unitary [hence the infinitesimal $\delta L$ obtained from Lie algebra is anti-Hermitian and has the basis $i\sigma^a$ ($a$=1,2,3)]. If the transformation operator $U$ characterizes Lorentz boost, for example, the Lorentz boost along $z$ direction [1]

$$U = \begin{bmatrix} U_2 & S_1 & 0 & S_3 \\ S_1^T & U_2 & -S_3 & 0 \\ 0 & 0 & U_1 & S_2 \\ 0 & 0 & S_2^T & U_1 \end{bmatrix}. \tag{B7}$$

Here, the blocks $U_1$ and $U_2$ are no longer rotation matrices, and off-diagonal blocks are present. There blocks are in the following forms

$$U_1 = \begin{bmatrix} \gamma & 0 \\ 0 & \gamma \end{bmatrix}, U_2 = \begin{bmatrix} \gamma & 0 & 0 \\ 0 & \gamma & 0 \\ 0 & 0 & 1 \end{bmatrix}, S_1 = \begin{bmatrix} 0 & \beta\gamma & 0 \\ -\beta\gamma & 0 & 0 \\ 0 & 0 & 0 \end{bmatrix},$$

$$S_2 = \begin{bmatrix} 0 & -\beta\gamma \\ \beta\gamma & 0 \end{bmatrix}, S_3 = \begin{bmatrix} 0 & \beta\gamma \\ -\beta\gamma & 0 \\ 0 & 0 \end{bmatrix}. \tag{B8}$$

where $\gamma = 1/\sqrt{1-\beta^2}$ is the Lorentz factor. It is shown that $U$ still has determinant 1 just as SO rotations, and the transformation process has a similar form $U^\dagger M U \psi'(x^\mu) = \omega U^\dagger N U \psi'(x^\mu)$, with $\psi'(x^\mu) = U^{-1}\psi(x^\mu)$. The transformed Hamiltonian is $H' = T^\dagger H T$ (with $T = N^{1/2} U N^{-1/2}$), but the eigenstate and the eigen frequency will be $\tilde{\psi}' = T^{-1}\tilde{\psi}$ and $\tilde{\omega} = \omega T^\dagger T \neq \omega$. Note that $TT^\dagger \neq 1$ for Lorentz boost, and now $T$ is non-unitary but has the determinant 1. For this case, the effective operator $L$ obtained from k.p theory is also non-unitary and has determinant 1, and thus the infinitesimal $\delta L$ is Hermitian and has the basis $\sigma^a$ ($a=1,2,3$).

### Appendix C: Deriving the spin connection

Here we provide a brief introduction on the derivation processes of spin connection. The relation between torsion two-form and spin connection one-form is given by Eq. (22). Firstly we take a look at the commutation relation

$$[E_a, E_b] = -\Omega_{ab}{}^c E_c, \tag{C1}$$

where $E_a$ is a contravariant vector expressed with the coframe field $E_a = e_a^\mu \partial_\mu$, with $\Omega_{ab}{}^c$ being the anholonomy coefficients and can be easily derived as

$$\Omega_{ab}{}^c = e_a^\mu e_b^\nu (\partial_\mu e_\nu^c - \partial_\nu e_\mu^c), \tag{C2}$$

the indices of which can be raised or lowered with metric tensors, and can also be transformed from Latin indices to Greek indices with frame or coframe fields (e.g. $\Omega_{ab}{}^c e_\mu^a e_\nu^b = \Omega_{\mu\nu}{}^c$, $\Omega_{ab}{}^c \eta_{cd} = \Omega_{ab}{}^d$). We find that the bracket in Eq. (C2) includes the total derivative on the frame field one form ($de^a$). Thus Eq. (C2) can be rewritten with Greek indices

$$\Omega_{\mu\nu\rho} = (\partial_\mu e_\nu^a - \partial_\nu e_\mu^a) e_{a\rho}. \tag{C3}$$

It can be easily demonstrated that the anholonomy coefficient is anti-symmetric about the first two indices. Equation (22) can be simplified as

$$T_{\mu\nu\rho} = \Omega_{\mu\nu\rho} + \omega_{\mu\rho\nu} - \omega_{\nu\rho\mu}. \tag{C4}$$

Treating the components of spin connection as unknowns, Eq. (C4) forms a set of linear equations, and can be solved analytically

$$\begin{aligned}
\omega_{\mu\nu\rho} &= \omega_{\mu\nu\rho}(e) + K_{\mu\nu\rho} \\
\omega_{\mu\nu\rho}(e) &= \frac{1}{2}(\Omega_{\mu\nu\rho} - \Omega_{\nu\rho\mu} + \Omega_{\rho\mu\nu}) = \omega_{\mu ab}(e) e_\nu^a e_\rho^b, \\
K_{\mu\nu\rho} &= -\frac{1}{2}(T_{\mu\nu\rho} - T_{\nu\rho\mu} + T_{\rho\mu\nu})
\end{aligned} \tag{C5}$$

where $\omega_{\mu\nu\rho}(e)$ is the torsion free spin connection (torsion free means $T_{\mu\nu\rho} = 0$), and $K_{\mu\nu\rho}$ is the contorsion tensor defined as the difference between $\omega_{\mu\nu\rho}$ and $\omega_{\mu\nu\rho}(e)$. One can easily find that the contorsion is also anti-symmetric about the last two indices, and the torsion comes from the asymmetry of the last two indices of affine connection $T_{\mu\nu}^\rho = \Gamma_{\mu\nu}^\rho - \Gamma_{\nu\mu}^\rho$. If there is no torsion, the contorsion is also zero, and the spin connection reduces to the torsion free spin connection. The torsion free spin connection [second equation in Eq. (C5)] is totally dependent on the anholonomy coefficients (or frame field). If the torsion term is present, the commutation between two covariant derivatives changes, for vector fields $[\nabla_\mu, \nabla_\nu] V^\tau = -T_{\mu\nu}^\rho \nabla_\rho V^\tau$, for Dirac spinors $[\nabla_\mu, \nabla_\nu] \phi = (1/4 R_{\mu\nu ab} \gamma^{ab} - T_{\mu\nu}^\rho \nabla_\rho) \phi$. These commutation relations can be demonstrated easily, and thus we will not introduce in detail here.

## Appendix D: Comparison with a magnetic field

Different from a gravity background, any spinor (such as Dirac, Weyl spinors) or scalar field in a magnetic field background has the local U(1) transformation $\phi' = \exp(i\alpha(x))\phi$, where $\alpha(x)$ is a scalar function that depends on spatial coordinates. The parallel transport of the spinor requires the partial derivative be replaced with the covariant derivative $\nabla_\mu = \partial_\mu + iA_\mu$, where $A_\mu$ is the U(1) gauge field, and the (synthetic) magnetic field comes from the commutation of covariant derives $[\nabla_\mu, \nabla_\nu] = i(\partial_\mu A_\nu - \partial_\nu A_\mu) = iB_{\mu\nu}$ (note that $B_{xy} = B_z$, $B_{yz} = B_x$ and $B_{zx} = B_y$). The artificial magnetic field can be realized by introducing inhomogeneous modulations to spatially control the position of Weyl degeneracy in *k* space, which was already reported in our previous work *(22)*. Non-Abelian gauge field can also be obtained in this way *(47)*. To solve the eigenstates, we square the Dirac operator

$$(g^{\mu\nu}\nabla_\mu\nabla_\nu - \frac{i}{2}B_{\mu\nu}\gamma^\nu\gamma^\mu)\phi = 0. \tag{D1}$$

Put the second term on the right and take the square on both sides, one obtain

$$(g^{\mu\nu}\nabla_\mu\nabla_\nu)^2\phi = -\frac{1}{4}(B_{\mu\nu}\gamma^\nu\gamma^\mu)^2\phi = -\frac{1}{8}B_{\mu\nu}B_{\rho\sigma}\{\gamma^\nu\gamma^\mu,\gamma^\sigma\gamma^\rho\}\phi. \tag{D2}$$

For magnetic field, $B_{\mu\nu}B_{\rho\sigma}$ has an interesting property,

$$B_{\mu\nu}B_{\rho\sigma} + B_{\nu\rho}B_{\mu\sigma} + B_{\rho\mu}B_{\nu\sigma} = 0, \quad B_{\mu\nu}B_{\rho\sigma} = B_{\rho\sigma}B_{\mu\nu}, \tag{D3}$$

with which one can easily derive

$$(g^{\mu\nu}\nabla_\mu\nabla_\nu)^2\phi = \frac{1}{2}g^{\mu\sigma}g^{\nu\rho}B_{\mu\nu}B_{\sigma\rho}\phi. \tag{D4}$$

Taking the square root, Eq. (D4) reads

$$g^{\mu\nu}\nabla_\mu\nabla_\nu\phi = \pm F\phi, \tag{D5}$$

with $F = \sqrt{g^{\mu\sigma}g^{\nu\rho}B_{\mu\nu}B_{\sigma\rho}}$. It is found that *F* is just part of the square root of background electromagnetic energy density $|\mathbf{E}|^2 + |\mathbf{B}|^2$, and if only magnetic field exsit, the square root of energy density reduces to *F*. If the magnetic field is uniformly distributed, Eq. (D5) can be solved analytically, giving the Landau levels. Considering a simple case that the magnetic field is along $y$ direction ($B_y = B_{zx}$, $[\nabla_x, \nabla_z] = -iB_{zx}$), for Weyl points $Q1$ and $Q1'$ in our inhomogeneous system, Eq. (D5) can be expanded as

$$\left((v_x^2 - w^2)\nabla_x^2 + v_y^2 \partial_y^2 + v_z^2 \nabla_z^2 - \partial_t^2 + 2w\nabla_x \partial_t\right)\phi = \pm\sqrt{\frac{g^{zz}g^{xx}B_{zx}B_{zx} + g^{xx}g^{zz}B_{xz}B_{xz}}{2}}\phi, \quad (D6)$$

and then simplified as

$$\left[(\sqrt{v_x^2 - w^2}\nabla_x + \frac{w}{\sqrt{v_x^2 - w^2}}\partial_t)^2 + v_y^2\partial_y^2 + v_z^2\nabla_z^2 - \frac{v_x^2}{v_x^2 - w^2}\partial_t^2\right]\phi = \pm\sqrt{v_x^2\gamma^{-2}v_z^2 B_{zx}^2}\phi. \quad (D7)$$

We can use the commutation of covariant derivative to define the ladder operators

$$a = \frac{1}{N}\left(\sqrt{v_x^2 - w^2}\nabla_x + \frac{w}{\sqrt{v_x^2 - w^2}}\partial_t - iv_z \nabla_z\right),$$

$$a^\dagger = \frac{1}{N}\left(\sqrt{v_x^2 - w^2}\nabla_x + \frac{w}{\sqrt{v_x^2 - w^2}}\partial_t + iv_z \nabla_z\right) \quad (D8)$$

and $[a, a^\dagger] = 1$ requires $N^2 = 2B_{zx}v_z\sqrt{v_x^2 - w^2}$. The harmonic form Eq. (D7) was rewritten as

$$\left[-(2\hat{n}+1)\sqrt{v_x^2\gamma^{-2}v_z^2 B_{zx}^2} + v_y^2\partial_y^2 - \frac{v_x^2}{v_x^2 - w^2}\partial_t^2\right]\phi = \pm\sqrt{v_x^2\gamma^{-2}v_z^2 B_{zx}^2}\phi \quad (D9)$$

and one can quickly derive the discrete Landau levels

$$\omega_n = \begin{cases} \pm\sqrt{v_y^2\kappa_y^2 + 2nv_x\gamma^{-1}v_z|B_{zx}|}\gamma^{-1}, & n \geq 1 \\ +/-v_y\kappa_y\gamma^{-1}, & +/- \text{ depends on } \text{sign}(v_xv_yv_zB_{zx}) \ n=0 \end{cases} \quad (D10)$$

with $\gamma = (1 - w^2/v_x^2)^{-1/2}$ being the Lorentz factor. The solution Eq. (D10) corresponds to the chiral magnetic effect, and experimental signature was observed in [22]. It is also noted that the Lorentz factor $\gamma$ should be real so that the Landau levels do not collapse, which requires $w \leq v_x$ [28]. The dispersions of Landau levels calculated by Eq. (D10) is shown in FIG. 7 (a)-(b).

If the magnetic field is not uniformly distributed, the harmonic form Eq. (D7) cannot be solved analytically and requires numerical tools. We set the magnetic field to be distributed as $B_{zx} = m_1 m_2 / (m_2 z + m_3)^2$ as an example. The corresponding U(1) gauge field is thus $A_x = -m_1 (m_2 z + m_3)^{-1}$. The dispersions of discrete energy levels are shown in FIG. 7 (c)-(d), which are obtained with PDE module of COMSOL. It is found that compared with Landau levels of a uniform magnetic field, the energy gaps between high order levels $n \geq 1$ are squeezed a lot in the dispersion diagram so that they are not conspicuous to the eye, but these levels are not degenerate. While the zeroth order level is almost identical with the case of uniform magnetic field. The Landau levels in dispersionless in the other direction independent of whether the magnetic field is uniformly or nonuniformly distributed. An axial electric field can generate a group velocity in this direction but will not change the dispersion profile.

For a gravitational field background, the Dirac operator can also be squared $\gamma^\mu \nabla_\mu (\gamma^\nu \nabla_\nu \phi) = 0$. Employing the relation of Clifford algebra (sentences following Eq. 6), this equation can be simplified as

$$(g^{\mu\nu} \nabla_\mu \nabla_\nu - \frac{1}{8} R_{\mu\nu ab} \gamma^\nu \gamma^\mu \gamma^{ab}) \phi = 0. \tag{D11}$$

To derive this equation, we need some commutation relations. We note that the frame field $e_a^\mu$ has two transformations with respect to $\mu$ and $a$, with the former being the general coordinate transformation and the latter being a local Lorentz transformation. It is noted that for Lorentzian vectors, the bases of Lorentz transformation $\Lambda_a^{\ b} = \exp(1/2 \lambda_{ab} E^{ab})$ are

$$E^{01} = \begin{bmatrix} 0 & 1 & 0 & 0 \\ 1 & 0 & 0 & 0 \\ 0 & 0 & 0 & 0 \\ 0 & 0 & 0 & 0 \end{bmatrix}, E^{02} = \begin{bmatrix} 0 & 0 & 1 & 0 \\ 0 & 0 & 0 & 0 \\ 1 & 0 & 0 & 0 \\ 0 & 0 & 0 & 0 \end{bmatrix}, E^{03} = \begin{bmatrix} 0 & 0 & 0 & 1 \\ 0 & 0 & 0 & 0 \\ 0 & 0 & 0 & 0 \\ 1 & 0 & 0 & 0 \end{bmatrix},$$

$$E^{12} = \begin{bmatrix} 0 & 0 & 0 & 0 \\ 0 & 0 & 1 & 0 \\ 0 & -1 & 0 & 0 \\ 0 & 0 & 0 & 0 \end{bmatrix}, E^{23} = \begin{bmatrix} 0 & 0 & 0 & 0 \\ 0 & 0 & 0 & 0 \\ 0 & 0 & 0 & 1 \\ 0 & 0 & -1 & 0 \end{bmatrix}, E^{31} = \begin{bmatrix} 0 & 0 & 0 & 0 \\ 0 & 0 & 0 & 1 \\ 0 & 0 & 0 & 0 \\ 0 & -1 & 0 & 0 \end{bmatrix} \quad (D12)$$

which is one-to-one correspondence to the basis $\gamma^{ab}$ (actually the bases for Eq. 13 is the same form as Eq. D12, but the indices are replaced by $\mu$ and $\nu$). Hence the covariant derivative on $e_a^\mu$ is $\nabla_\mu e_a^\nu = \partial_\mu e_a^\nu + \omega_{\mu a b} e^{b\nu} - \Gamma_{\mu\rho}^\nu e_a^\rho$, which is demonstrated to be zero, and thus $[\nabla_\mu, e_a^\nu] = 0$. The coordinate based Dirac matrix $\gamma_\mu$ is a vector formed bi-spinor, which means that it has four components ($\mu = t, x, y, z$), and each component is a 4 by 4 matrix that can regarded as the direct product Dirac spinors $\phi\bar{\phi}$, with $\bar{\phi} = \phi^\dagger i\gamma^0$ being the Dirac adjoint. Hence, except for the transformation characterized by Eq. (12-13), there is another transformation, with the infinitesimal of transformation operation $\delta[\phi\bar{\phi}] = \delta\phi\bar{\phi} + \phi\delta\bar{\phi} = -1/4\lambda_{ab}[\gamma^{ab}, \phi\bar{\phi}]$. Hence, the covariant derivative is $\nabla_\mu \gamma_\nu = \partial_\mu \gamma_\nu + 1/4\omega_\mu^{ab}[\gamma_{ab}, \gamma_\nu] - \Gamma_{\mu\nu}^\rho \gamma_\rho$, which can also be demonstrated to be zero ($[\nabla_\mu, \gamma_\nu] = 0$). Besides, the covariant derivative also commutes with the matric tensors (i.e. $[\nabla_\mu, \eta_{ab}] = 0$ and $[\nabla_\mu, g_{\nu\rho}] = 0$), and we will not discuss in detail here. The covariant derivative obeys the Leibniz product rule [2], for which we have an example $\nabla_\mu \gamma^\nu = \nabla_\mu (\gamma_\rho g^{\nu\rho}) = (\nabla_\mu \gamma_\rho) g^{\nu\rho} + \gamma_\rho \nabla_\mu g^{\nu\rho} = 0$. With these commutation relations, Eq. (D11) can be simplified [2]

$$(g^{\mu\nu}\nabla_\mu \nabla_\nu - \frac{1}{4}R)\phi = 0. \quad (D13)$$

with $R = R_{\mu\nu\rho\sigma}g^{\mu\rho}g^{\nu\sigma} = R_{\nu\sigma}g^{\nu\sigma}$ being the scalar Ricci curvature, and $R_{\nu\sigma} = R_{\mu\nu\rho\sigma}g^{\mu\rho}$ the curvature tensor. In deriving Eq. (D13), the Bianchi identity and permutation symmetry for torsion free gravitational field are used

$$R_{\mu\nu\rho\sigma} + R_{\nu\rho\mu\sigma} + R_{\rho\mu\nu\sigma} = 0, \ R_{\mu\nu\rho\sigma} = R_{\rho\sigma\mu\nu}, \tag{D14}$$

which is very similar to Eq. (D14). We found that the scalar $R$ in Eq. (D14) acts as the similar role to $F$ in Eq. (D14), but $R$ cannot be imaginary (not from a square root just like $F$), and hence the discrete energy levels do not collapse, which is an essential difference.

## Appendix E: Can the spin connection term be zero?

Previously we have shown that in the inhomogeneous system with spatially controlled non-locality, the group velocity near the Weyl cone depends on spatial position, and thus, in the Weyl and Dirac equations, the partial derivative is replaced with covariant derivative to ensure the parallel transport of the pseudo-spinor. The non-Abelian gauge field $(1/4\,\omega_{\mu ab}\gamma^{ab})$ in the covariant derivative is not obvious in the *k.p* perturbation theory, and thus someone may ask, if the spin connection is ignored, is the Weyl or Dirac equation still valid? In the following we provide an example to demonstrate that spin connection is essential to obtain physically correct energy levels and thus cannot be omitted.

To simplify the derivation and provide an analytical result, we set the tilt velocity to be zero ($w=0$), and set the group velocity $v_x$ to be linearly dependent on $z$ ($v_x = mz + m'$, with $m$ and $m'$ being constants). The coframe field is thus

$$e_0^\mu = (1,0,0,0),\ e_1^\mu = (0,v_x,0,0),\ e_2^\mu = (0,0,v_y,0),\ e_3^\mu = (0,0,0,v_z). \tag{E1}$$

If the spin connection is zero, from Eq. (23) one can easily find that the torsion

$$T^a_{\mu\nu} = \partial_\mu e^a_\nu - \partial_\nu e^a_\mu. \tag{E2}$$

To solve the energy levels from Dirac equation, we need to square the Dirac equation

$$e^\mu_a \gamma^a \partial_\mu (e^\nu_b \gamma^b \partial_\nu \phi) = 0. \tag{E3}$$

One can easily find the commutation relation

$$[e^\mu_a \partial_\mu, e^\nu_a \partial_\nu] = [E_a, E_b] = -e^\mu_a e^\nu_b (\partial_\mu e^c_\nu - \partial_\nu e^c_\mu) E_c = -e^\mu_a e^\nu_b T^c_{\mu\nu} E_c. \tag{E4}$$

Thus Eq. (E3) reduces to

$$g^{\mu\nu} \partial_\mu \partial_\nu \phi - \frac{1}{2} \gamma^a \gamma^b e^\mu_a e^\nu_b T^c_{\mu\nu} E_c \phi = 0. \tag{E5}$$

in which the Clifford (or Dirac) algebra is used. For the current case, Eq. (E5) can be expanded as

$$(-\partial_t^2 + v_x^2 \partial_x^2 + v_y^2 \partial_y^2 + v_z^2 \partial_z^2)\phi - \frac{1}{2}(\gamma^1 \gamma^3 v_x v_z T^1_{xz} v_x + \gamma^3 \gamma^1 v_z v_x T^1_{zx} v_x)\partial_x \phi = 0. \tag{E6}$$

The only nonzero term of torsion is $T^1_{xz} = -T^1_{zx} = -\partial_z v_x^{-1} = v'_x v_x^{-2}$, and thus we have

$$(-\partial_t^2 + v_x^2 \partial_x^2 + v_y^2 \partial_y^2 + v_z^2 \partial_z^2)\phi = \gamma^{13} v_z v_x^2 T^1_{xz} \partial_x \phi. \tag{E7}$$

We can use the commutation $[E_1, E_3] = -v_z v_x^2 T^1_{xz} \partial_x = -iv_z v'_x k_x$ (being independent of spatial coordinate) to define the ladder operators,

$$a = \frac{1}{N}(E_1 + iE_3), \quad a^\dagger = \frac{1}{N}(E_1 - iE_3). \tag{E8}$$

The definition of ladder operators requires $[a, a^\dagger] = 1$, with which we have $N^2 = -2v_z v'_x k_x$. The number operator is thus

$$\hat{n} = a^\dagger a = \frac{1}{N^2}(E_1^2 + i[E_1, E_3] + E_3^2), \tag{E9}$$

and Eq. (E7) can be rewritten in terms of the number operator

$$(\omega^2 - v_y^2 \kappa_y^2 + N^2 \hat{n} - v_z v'_x \kappa_x)\phi = i\gamma^{13} v_z v'_x \kappa_x \phi \tag{E10}$$

One can then derive the energy levels from Eq. (E10),

$$\omega_n = \begin{cases} \pm\sqrt{v_y^2\kappa_y^2 + 2nv_z m\kappa_x}, & n \geq 1 \\ +/-v_y\kappa_y, \ +/- \text{ depends on chiral and sign}(m) & n=0 \end{cases} \quad \text{(E11)}$$

Since for high order energy levels ($n \geq 1$), the term $2nv_z v'_x \kappa_x$ can be positive or negative, and the term $v_y^2\kappa_y^2$ varies from zero to infinity, it is possible that the sign in the square root is negative, and the corresponding eigen-frequency is imaginary. *As shown in FIG. 8, exceptional points appear in the energy bands for high order levels, which is unphysical since the inhomogeneous modulation cannot change the Hermiticity of the system. Thus we can conclude that in the inhomogeneous system with spatially controlled Weyl or Dirac cone dispersions, partial derivatives must be replaced with covariant derivatives, in which a non-Abelian gauge field (with spin connection) must be retained to ensure the parallel transport of the pseudo-spinor.*

**Appendix F: Frame fields for other Weyl points.**

We previously provided detailed analysis of Weyl point *Q1* and its partner *Q1'*. The frame field, as well as the metric tensor and its inverse can be easily obtained from Eq. (9) by using the orthogonal relation introduced in the main text, and are expressed as

$$e_\mu^0 = (1,0,0,0), \ e_\mu^1 = (wv_x^{-1}, v_x^{-1}, 0, 0), \ e_\mu^2 = (0, 0, v_y^{-1}, 0), \ e_\mu^3 = (0,0,0,v_z^{-1}), \quad \text{(F1)}$$

$$g_{\mu\nu} = \begin{bmatrix} -(1-w^2/v_x^2) & w/v_x^2 & 0 & 0 \\ w/v_x^2 & v_x^{-2} & 0 & 0 \\ 0 & 0 & v_y^{-2} & 0 \\ 0 & 0 & 0 & v_z^{-2} \end{bmatrix}, \ g^{\mu\nu} = \begin{bmatrix} -1 & w & 0 & 0 \\ w & v_x^2(1-w^2/v_x^2) & 0 & 0 \\ 0 & 0 & v_y^2 & 0 \\ 0 & 0 & 0 & v_z^2 \end{bmatrix}. \quad \text{(F2)}$$

which can be used to calculate the Levi-Civita connection Eqs. (24-25). The chiralities of the four Weyl points are positive for *Q1* and *Q3*, and negative for *Q2* and *Q4*. Thus the chiralities for their four partners *Q1-Q4* are just the inverse of each other. All the four Weyl cones are tilted towards the center in **k** space. Since the analysis of other Weyl points are quite similar to *Q1* and *Q1'*, we just need to provide the frame field, so that the spin connection and Riemannian curvatures can be obtained as an analogue with previous discussions. Since all the velocity components $w$, $v_x$, $v_y$, $v_z$ are positive, Weyl points *Q3* and *Q3'* together can be described with the Dirac equation

$$(\gamma^0 \partial_t + w\gamma^0 \partial_x - v_x \gamma^1 \partial_x - v_y \gamma^2 \partial_y + v_z \gamma^3 \partial_z)\phi = 0. \tag{F3}$$

This equation is obtained from *k.p* perturbation theory, it is shown that the signs of $w$, $v_x$ and $v_y$ are inversed compared with *Q1* and *Q1'* [see Eq. (9)]. The coframe and frame fields can thus be obtained as

$$\begin{aligned} e_0^\mu &= (1, w, 0, 0),\ e_1^\mu = (0, -v_x, 0, 0),\ e_2^\mu = (0, 0, -v_y, 0),\ e_3^\mu = (0, 0, 0, v_z) \\ e_\mu^0 &= (1, 0, 0, 0),\ e_\mu^1 = (-wv_x^{-1}, -v_x^{-1}, 0, 0),\ e_\mu^2 = (0, 0, -v_y^{-1}, 0),\ e_\mu^3 = (0, 0, 0, v_z^{-1}) \end{aligned}. \tag{F4}$$

For *Q2* and *Q2'*, the Dirac equation obtained from *k.p* perturbation is

$$(\gamma^0 \partial_t - w\gamma^0 \partial_y - v_y \gamma^1 \partial_x + v_x \gamma^2 \partial_y + v_z \gamma^3 \partial_z)\phi = 0. \tag{F5}$$

and the corresponding frame fields

$$\begin{aligned} e_0^\mu &= (1, 0, -w, 0),\ e_1^\mu = (0, -v_y, 0, 0),\ e_2^\mu = (0, 0, v_x, 0),\ e_3^\mu = (0, 0, 0, v_z) \\ e_\mu^0 &= (1, 0, 0, 0),\ e_\mu^1 = (0, -v_y^{-1}, 0, 0),\ e_\mu^2 = (-wv_x^{-1}, 0, v_x^{-1}, 0),\ e_\mu^3 = (0, 0, 0, v_z^{-1}) \end{aligned}. \tag{F6}$$

For *Q4* and *Q4'*, we have

$$(\gamma^0 \partial_t + w\gamma^0 \partial_y + v_y \gamma^1 \partial_x - v_x \gamma^2 \partial_y + v_z \gamma^3 \partial_z)\phi = 0. \tag{F7}$$

$$\begin{aligned} e_0^\mu &= (1, 0, w, 0),\ e_1^\mu = (0, v_y, 0, 0),\ e_2^\mu = (0, 0, -v_x, 0),\ e_3^\mu = (0, 0, 0, v_z) \\ e_\mu^0 &= (1, 0, 0, 0),\ e_\mu^1 = (0, v_y^{-1}, 0, 0),\ e_\mu^2 = (wv_x^{-1}, 0, -v_x^{-1}, 0),\ e_\mu^3 = (0, 0, 0, v_z^{-1}) \end{aligned}. \tag{F8}$$

Under inhomogeneous perturbations, the velocity components $w$ and $v_x$ are still linear functions of $z$, the connection and curvature can be obtained analytically just like the deriving process of $Q1$ and we will not repeat it here. The discrete energy levels are numerically obtained in the main text.

**Appendix G: Impact of loss on the presence of Weyl point and estimate of sample size**

The impact of loss for the periodic system can be numerically investigated to check if the Weyl points are still present in case that the loss is introduced. In general, the loss may come from the metallic wires or the dielectrics in the structure (purple and green blocks FIG. 9a). If the loss come from the wire, it introduces an imaginary part of $\omega_0$ in Eq. (1), which will result in the expanding of the Weyl point to an exceptional ring [46]. However, in the microwave region, the skin depth of electromagnetic field is very small, the metallic wires are well represented by ideal PEC, and the loss mainly comes from the dielectrics. In the following we provide some numerical results to show the impact of the loss from the dielectrics.

The introduction of loss can be easily realized by adding an imaginary part of the dielectric constants $\varepsilon = \varepsilon' + i\varepsilon''$ of the dielectric blocks (purple and green) in FIG. 9(a). In the absence of loss, the frequencies of eigen states forming the Weyl crossing are real (taking Weyl point $Q1$ as an example, FIG. 9a). After loss is introduced ($\varepsilon''$ is nonzero), the eigen frequencies become complex, and the imaginary parts has the same dispersion profile as the real parts (FIG. 9 b1-b2). Even though the eigen frequencies become complex, the Weyl point still exists, without changing the shape of the Weyl cone. Increasing the loss (increasing the value of $\varepsilon''$) can lead to the larger value of the imaginary parts of eigen frequencies (FIG. 9 b2-f2), but the real parts almost stay unchanged (FIG. 9 b1-f1). We note that the PEC excludes the field and all the electric fields are in the dielectrics. As the imaginary part of $\varepsilon$ is the same everywhere inside the dielectric, we do not have a "differential loss parameter" and hence the uniform loss parameter does not induce exceptional points. As such, the loss from the dielectric does not change the observables proposed in this work.

We can give an estimate on the size of the non-periodic system by using the inequality $(1/v)(\Delta v/\Delta L) \ll 1/\lambda$, where $v$ is the group velocity of the longitudinal mode near the Weyl cone (i.e. $v_{gx,-}$ in the main text), $\Delta L$ is the size of the metamaterial in the non-periodic direction and $\lambda$ is the wavelength at the Weyl cone (around 3mm for the designed sample in FIG. 5). The group velocity $v$ is at the order of $10^8$m/s, and the variation of the group velocity $\Delta v$ is around $0.5 \times 10^8$m/s. If we make the right hand side 3 to 4 times of the left hand side (can be treated as much larger, i.e. "$\ll$" in the equality), the size of the sample in the non-periodic direction should be around 50cm to 70cm. The size of the unit cell ($h_1+h_2$ in FIG. 5a) is 6mm in average, and hence approximately 100 unit cells are required in the non-periodic direction. In the other periodic directions, 30 to 40 unit cells are required to retain the periodic property. Referring to a prior experimental work investigating synthetic magnetic field in graded Weyl metamaterials [22]. In that work, the signature of chiral magnetic effect in a spatially gradient metamaterial was observed in a metamaterial composing of 101 unit cells in the non-periodic direction, and 101×40 unit cells in the periodic directions. It is found that the number of unit cells is very close to our estimation and also shows that our system is experimentally realizable as similar-sized sample has been made before.

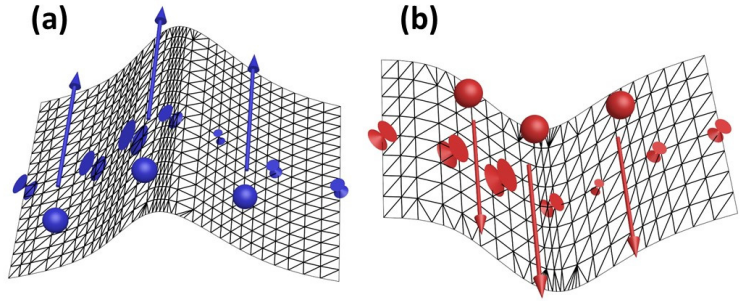

FIG. 1. Chiral transport of pseudo-spinors in Weyl metamaterial with synthetic gravitational field. (a) A curved space-time generated by a *Weyl cone* (blue cones) varying gradually from Type-II to Type-I from left to right, with the pseudo-spinor at chiral zeroth order mode propagating positively. (b) Reversing the tilt direction of the *Weyl cone* (red cones) leads to a flipped curved space-time, with the pseudo-spinor propagating negatively.

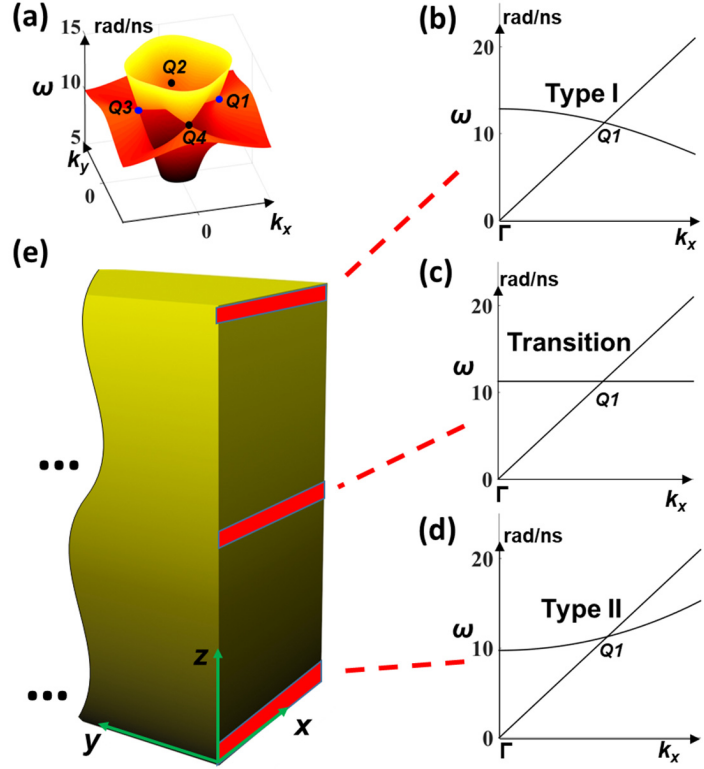

FIG. 2. Design of inhomogeneous metamaterial generating synthetic gravitational field. (a) Dispersion diagram at the $k_z=0$ plane [we set $f_1=3\times10^{20}$(rad/s)$^2$, $f_2=4$ in Eq. (1)]. There are four Weyl points (labelled by $Q1$ to $Q4$. $Q1$, $Q3$: positive chirality; $Q2$, $Q4$: negative chirality) located on the $k_x$ and $k_y$ axis. (B)-(D). Dispersion diagram for Weyl point $Q1$ along $+k_x$ direction ($k_y=k_z=0$), the Weyl cone achieves the transition from Type-I to Type-II by tuning the resonance frequency $\omega_0$ as a function of **k**, $\omega_0 = p(k_x^2+k_y^2)+q$. (b)-(d) correspond to $p<0$, $p=0$ and $p>0$, respectively. $q$ is elaborately tuned for any $p$ so that the Weyl frequency and position in $k$ space do not vary. (e) Design of inhomogeneous Weyl metamaterial generating curved geometry for pseudo-spinor. Non-locality $\omega_0(k_x,k_y)$ is spatially controlled ($p$ and $q$ vary in space), and gradient is along $z$ direction (homogeneous in $x$ and $y$ direction), realizing the linear dependence of $w$ and $v_x$ on $z$.

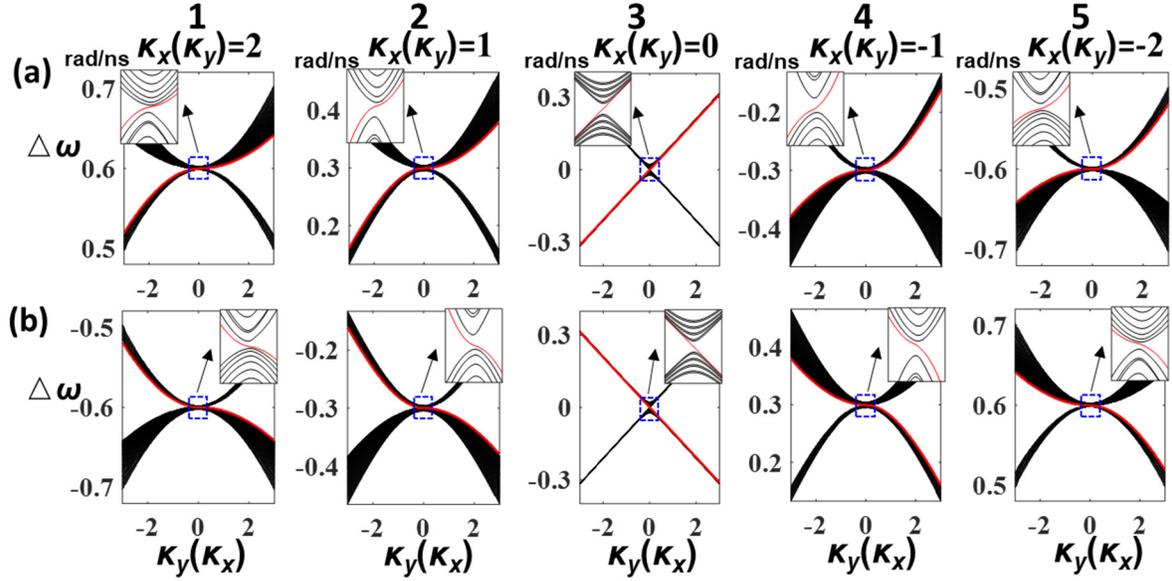

FIG. 3. Plots of eigen frequency obtained by solving covariant Weyl equation. The eigen-frequency, $\Delta\omega = \omega - \omega_W$ being the difference of frequency to Weyl frequency $\omega_W = 11.27$ rad/ns is plotted as functions of $\kappa_x$ and $\kappa_y$ (unit: rad/m). (a) corresponds to Weyl points $Q1$, and (b) corresponds $Q3$. For $Q2$ and $Q4$, the dispersion relations can be obtained by exchanging $\kappa_x$ and $\kappa_y$ in Row (a) and Row (a) (as indicated by the brackets), repectively. Panels 1-5 correspond to different $\kappa_x$ [for $Q1$ and $Q3$] or $\kappa_y$ [for $Q2$ and $Q4$] planes [$\kappa_x(\kappa_y) = 2, 1, 0, -1, -2$]. The red curves represent the zeroth order energy level, and the black curves represent high order energy levels. The insets in are zoomed in figures showing the quantization of energy levels.

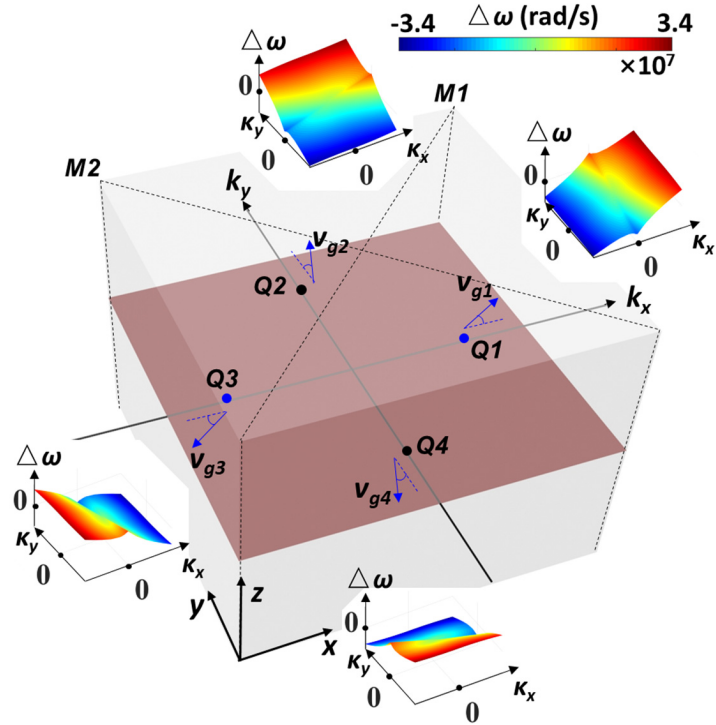

FIG. 4. Transport properties of chiral zeroth eigen modes. $v_{g1} \sim v_{g4}$ represent group velocities of zeroth mode for *Q1* to *Q4*. The four insets show the 2D dispersion of zeroth mode ($\triangle \omega$ plotted as a function of $\kappa_x$ and $\kappa_y$) for *Q1-Q4*, respectively, with colour scales showing the equi-frequency contours. *M1* and *M2* indicate two mirror planes (as enclosed by dashed lines) in the inhomogeneous system.

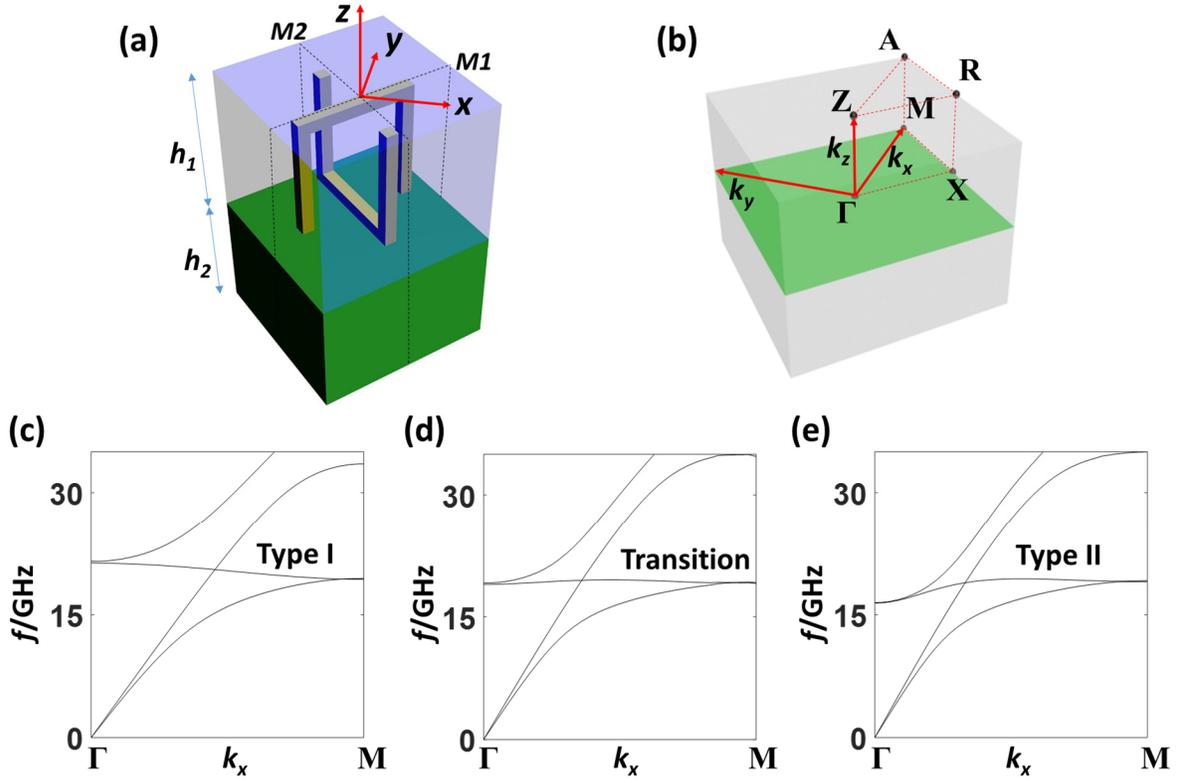

FIG. 5. Unit cell of metamaterial and band structures. (a) Structure of a unit cell of a Weyl metamaterial for realizing the constitutive parameters in Eq. (1). It has two dielectric layers with dielectric constants $\varepsilon_1$ and $\varepsilon_2$ (we set $\varepsilon_1=\varepsilon_2$) and the thicknesses are $h_1$ and $h_2$, respectively. Two orthogonal metallic wires with square cross sections forming resonant circuits are placed inside the first layer. The space inversion symmetry is broken, and there are two mirror symmetry plans (marked by *M1* and *M2*) with respect to *x* and *y*, consistent with the constitutive parameters of the metamaterial. (b) Brillouin zone of the structure. (c)-(e) Band structure on line Γ-M, corresponding to Type I Weyl cone, horizon between Type I and Type II and Type II Weyl cone respectively. The transition from Type I to Type II Weyl cone is achieved by tuning $h_2$ in (A). Results are obtained with CST microwave studio.

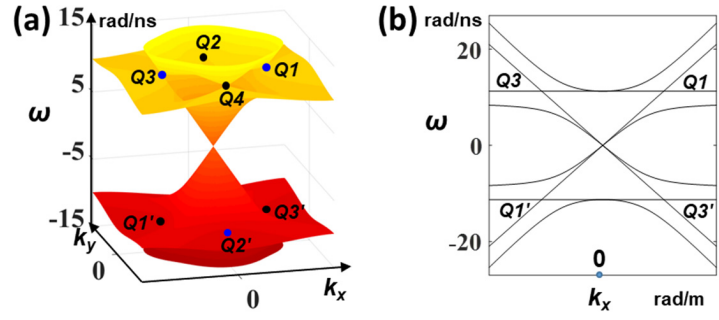

FIG. 6. Dispersion of energy bands forming the Weyl degeneracies. (a) $Q1$-$Q4$ are in the positive frequency region, and $Q1'$-$Q4'$ are in the negative frequency region. $Q1$ and $Q1'$ have opposite **k** vectors and opposite chiralities, as well as $Q2$ and $Q2'$, $Q3$ and $Q3'$, $Q4$ and $Q4'$. (b) energy band dispersion on $k_y=k_z=0$ line (i.e. $k_x$ axis). Eight energy bands are plotted in the figure, with four bands in the positive frequency region and four in the negative region.

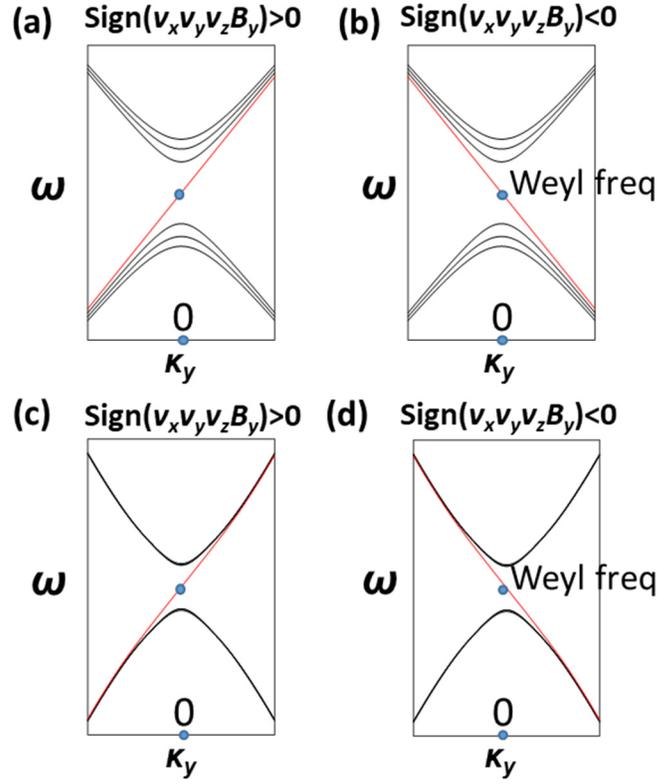

FIG. 7. Dispersion of Landau levels ($n = 0,1,2,3$). (a) and (b) for homogeneous magnetic field, (c) and (d) for inhomogeneous magnetic field. The red lines mark the zeroth order level and black curves represent higher order levels. (a) and (b) correspond to energy dispersions with different signs of chirality and magnetic field [i.e. $\text{Sign}(v_x v_y v_z B_y) > 0$ for (a) and (c), $\text{Sign}(v_x v_y v_z B_y) < 0$ for (b) and (d)], showing different group velocity directions of zeroth level. The gaps between high order Landau levels are squeezed substantially for inhomogeneous magnetic field. The magnetic field is distributed as $B_y = m_1 m_2 / (m_2 z + m_3)^2$, and the results are numerically obtained with the PDE module of COMSOL.

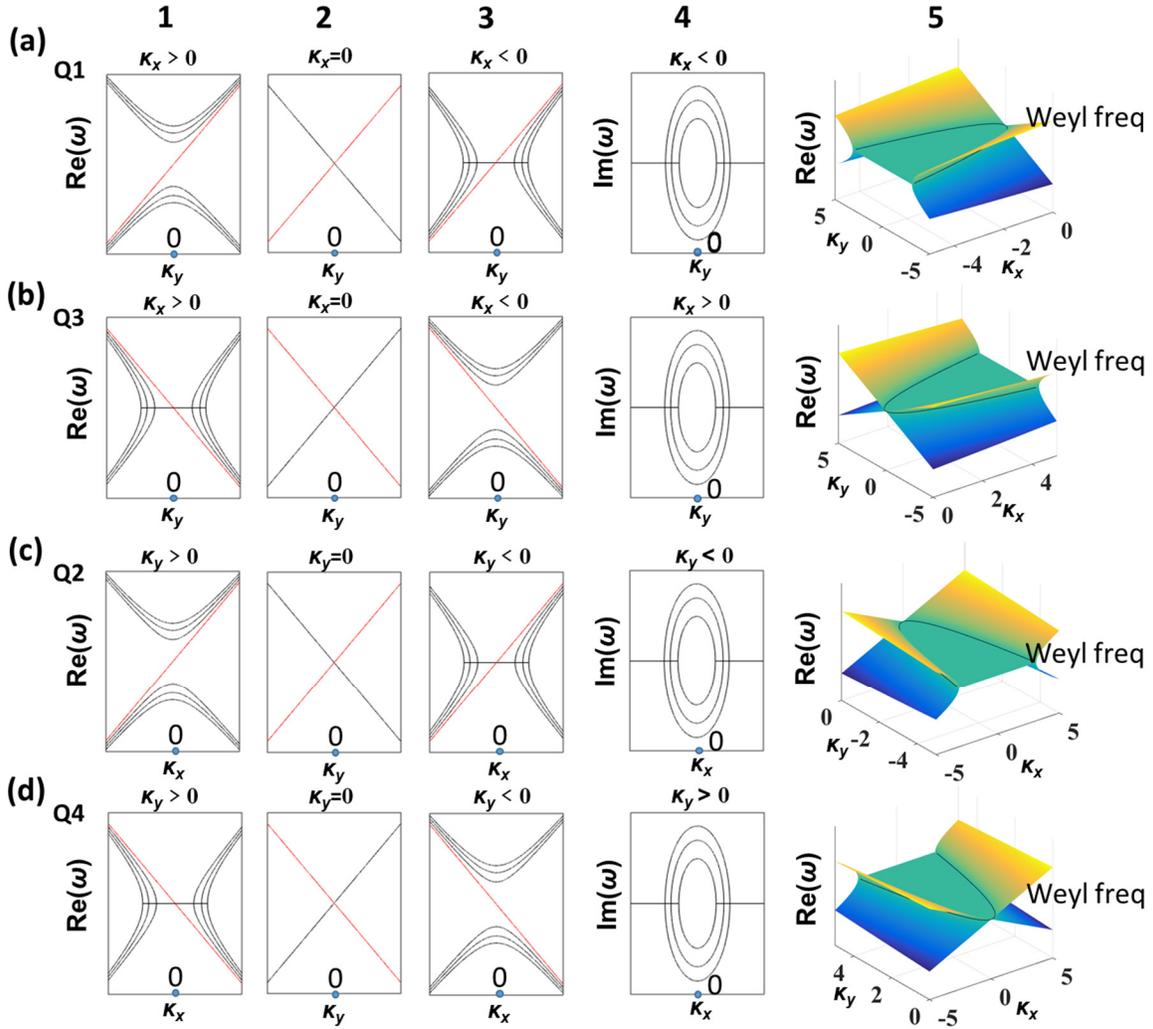

FIG. 8. Dispersion of energy levels calculated with Eq. (E11). (a)-(d) corresponds to Weyl points *Q1-Q4*, respectively. Panels 1-3 show the real part of eigen frequency of energy levels at different ( $>$, $=$ and $<0$) $\kappa_x$ (for *Q1* and *Q3*) planes or $\kappa_y$ (for *Q2* and *Q4*) planes. Panel 4 shows the imaginary part of eigen frequency at an specific plane as labelled in the figure. The zeroth energy level is displayed with red lines and other higher energy levels are shown with black lines in Panels 1-4. Panel 5 show the 2D (in $k_x k_y$ plane) energy dispersion of the two [+ and – in Eq. (E11)] first order energy levels, and the exceptional line is labelled with the black curves.

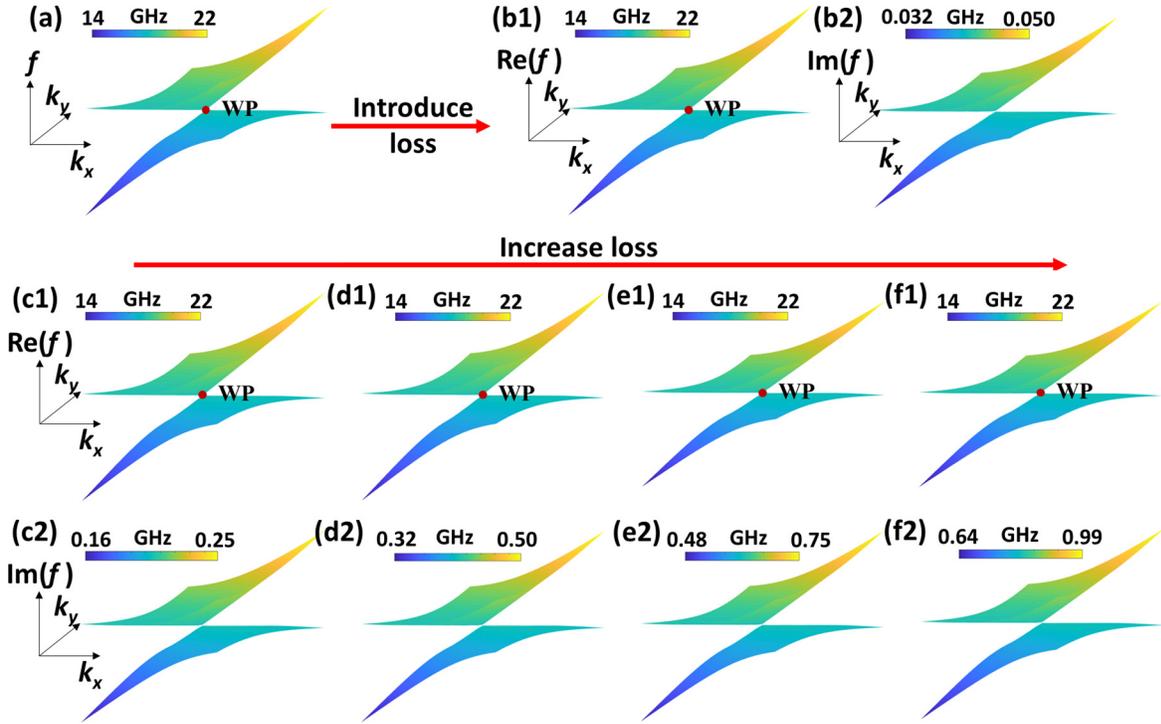

FIG. 9. Impact of loss on the presence of Weyl point. (a) in the absence of loss (b-f) introduce and increase loss in the dielectric part. (b1-f1) real part of eigen frequency. (b2-f2) imaginary part of eigen frequency. The loss is introduced by adding an imaginary part in the dielectric constant $\varepsilon = \varepsilon' + i\varepsilon''$ of the dielectrics in the structure FIG. 5(a) (purple and green blocks). Results are calculated with COMSOL RF module. (b-f) correspond to $\varepsilon'' = 0.001, 0.05, 0.1, 0.15, 0.2$, respectively. WP: Weyl point.